\documentclass[11pt,a4paper]{article}

\usepackage{amssymb}
\usepackage[dvips]{graphicx}
\usepackage{bm}

\begin{document}

\title{The higher covariant derivative regularization as a tool for revealing the structure of quantum corrections in supersymmetric gauge theories.}

\author{
K.V.Stepanyantz\\
{\small{\em Moscow State University,}}\\
{\small{\em Faculty of Physics, Department of Theoretical Physics,}}\\
{\small{\em 119991, Moscow, Russia}}}

\maketitle

\begin{abstract}
We discuss why the Slavnov higher covariant derivative regularization appeared to be an excellent instrument for investigating quantum corrections in supersymmetric gauge theories. For example, it allowed to demonstrate that the $\beta$-function in these theories is given by integrals of double total derivatives and to construct the NSVZ renormalization prescription in all loops. It was also used for deriving the non-renormalization theorem for the triple gauge-ghost vertices. With the help of this theorem the exact NSVZ $\beta$-function was rewritten in a new form, which revealed its perturbative origin. Moreover, in the case of using the higher covariant derivative regularization it is possible to construct a method for obtaining the $\beta$-function of ${\cal N}=1$ supersymmetric gauge theories, which simplifies the calculations in a great extent. This method is illustrated by an explicit two-loop calculation made in the general $\xi$-gauge.
\end{abstract}

\unitlength=1cm

\unitlength=1cm

%%%%%%%%%%%%%%%%%%%%%%%%%%%%%%%%%%%

\vspace*{-13.9cm}

\begin{flushright}
{\it Dedicated to the 80-th anniversary\\ of Prof. A.A.Slavnov}
\end{flushright}

\vspace*{12.2cm}

\section{Introduction}
\hspace*{\parindent}

Investigation of quantum corrections in various field theory models plays an important role for understanding nature. For instance, the comparison of the experimental values for the electron and muon anomalous magnetic moments with the theoretical predictions unambiguously indicates that nature should be described by quantum field theory \cite{Peskin:1995ev}. The renormalization group behaviour of the running coupling constants in the Standard model and its extensions can be interpreted as an indirect evidence of supersymmetry and Grand Unification \cite{Mohapatra:1986uf}. Certainly, there are also a large number of other examples. However, it is well known that usually quantum corrections are divergent in the ultraviolet region, so that for calculating them one should use a regularization. Although the most popular method for making perturbative calculations is dimensional regularization \cite{tHooft:1972tcz,Bollini:1972ui,Ashmore:1972uj,Cicuta:1972jf}, for supersymmetric theories it is very inconvenient because of the manifest breaking of supersymmetry \cite{Delbourgo:1974az}. Its special modification, called dimensional reduction \cite{Siegel:1979wq}, appears to be mathematically inconsistent \cite{Siegel:1980qs} and can also break supersymmetry in very higher orders \cite{Avdeev:1981vf,Avdeev:1982xy}. However, for regularizing supersymmetric theories one can use generalizations of the higher covariant derivative regularization proposed by A.A.Slavov in Refs. \cite{Slavnov:1971aw,Slavnov:1972sq}. An evident advantage of this regularization is that it is formulated in integer space-time dimensions. Moreover, in the supersymmetric case it can be consistently formulated in terms of ${\cal N}=1$ superfields \cite{Krivoshchekov:1978xg,West:1985jx}. It is also possible to construct ${\cal N}=2$ supersymmetric higher derivative regulators \cite{Buchbinder:2014wra}, but for theories with extended supersymmetry the version formulated in ${\cal N}=2$ harmonic superspace \cite{Galperin:1984av,Galperin:2001uw} is the most preferable. It has been constructed in Ref. \cite{Buchbinder:2015eva} and allows to calculate quantum corrections in a manifestly ${\cal N}=2$ supersymmetric way in all orders. Using this regularization one can derive the non-renormalization theorems for theories with extended supersymmetry \cite{Grisaru:1982zh,Mandelstam:1982cb,Brink:1982pd,Howe:1983sr} in a simple and elegant way \cite{Buchbinder:2015eva,Buchbinder:1997ib}.

The main idea of the higher covariant derivative regularization is to add a term with a large degree of the covariant derivatives to the classical action. This allows to remove all divergences beyond the one-loop approximation \cite{Faddeev:1980be}. For regularizing the remaining one-loop divergences one has to insert into the generating functional the Pauli--Villars determinants \cite{Slavnov:1977zf}. The presence of higher derivatives in the action essentially complicates explicit calculations of quantum corrections. For a long time this was a main obstacle for using this regularization. However, the calculations made in supersymmetric theories during the last decades demonstrated that the higher derivative regularization reveals the structure of quantum corrections in these theories and allows to solve the long-standing problems of deriving the NSVZ equation and constructing the NSVZ scheme. In this paper we will briefly review these issues and illustrate the results by an explicit two-loop calculation.

\section{The supersymmetric version of the higher covariant derivative regularization}
\hspace*{\parindent}\label{Section_Higher_ovariant_Derivatives}

It is convenient to formulate ${\cal N}=1$ supersymmetric theories in superspace, because in this case supersymmetry turns out to be a manifest symmetry. In this case a general massless renormalizable ${\cal N}=1$ supersymmetric gauge theory with a simple gauge group $G$ is described by the action

\begin{eqnarray}\label{Superfield_Action}
&& S_{\mbox{\scriptsize classical}} = \frac{1}{2e^2} \mbox{Re}\,\mbox{tr} \int d^4x\, d^2\theta\, W^a W_a + \frac{1}{4} \int d^4x\, d^4\theta\, \phi^{*i} (e^{2V})_i{}^j \phi_j\qquad\nonumber\\
&&\qquad\qquad\qquad\qquad\qquad\qquad\qquad\qquad\quad + \Big(\, \frac{1}{6} \lambda^{ijk} \int d^4x\, d^2\theta\, \phi_i \phi_j \phi_k + \mbox{c.c.}\Big),\qquad
\end{eqnarray}

\noindent
where $V$ is the gauge superfield, and $\phi_i$ are the chiral matter superfields. The supersymmetric gauge superfield strength $W_a = \bar D^2 (e^{-2V} D_a e^{2V})/8$ is a chiral superfield which transforms as a right spinor under the Lorentz group.

To construct the corresponding quantum theory, it is convenient to use the background field method \cite{DeWitt:1965jb,Abbott:1980hw,Abbott:1981ke} in the supersymmetric version \cite{Grisaru:1982zh,Gates:1983nr}. Moreover, it is necessary to take into account the necessity of the non-linear renormalization of the quantum gauge superfield, see the general consideration in Refs. \cite{Piguet:1981fb,Piguet:1981hh,Tyutin:1983rg} and explicit calculations of Refs. \cite{Juer:1982fb,Juer:1982mp,Kazantsev:2018kjx}. This can be done by making the substitution

\begin{equation}
e^{2V} \to e^{2{\cal F}(V)} e^{2\bm{V}},
\end{equation}

\noindent
where the function ${\cal F}(V)$ contains an infinite set of parameters needed for making the nonlinear renormalization, and $\bm{V}$ is the background gauge superfield. Note that in this case the quantum gauge superfield satisfies the constrain $V^+ = e^{-2\bm{V}} V e^{2\bm{V}}$.

To introduce the higher covariant derivative regularization we first modify the action in such a way that the regularized action contains higher degrees of the supersymmetric covariant derivatives,

\begin{eqnarray}\label{Action_With_HD_Regularization}
&& S_{\mbox{\scriptsize reg}} = \frac{1}{2 e_0^2}\,\mbox{Re}\, \mbox{tr} \int d^4x\, d^2\theta\, W^a \Big(e^{-2\bm{V}} e^{-2{\cal F}(V)}\Big)_{Adj} R\Big(-\frac{\bar\nabla^2 \nabla^2}{16\Lambda^2}\Big)_{Adj}\Big(e^{2{\cal F}(V)}e^{2\bm{V}}\Big)_{Adj} W_a \qquad\nonumber\\
&& + \frac{1}{4} \int d^4x\,d^4\theta\, \phi^{*i} \Big(F\Big(-\frac{\bar\nabla^2 \nabla^2}{16\Lambda^2}\Big) e^{2{\cal F}(V)}e^{2\bm{V}}\Big)_i{}^j \phi_j
+ \Big(\, \frac{1}{6} \lambda_0^{ijk} \int d^4x\,d^2\theta\, \phi_i \phi_j \phi_k + \mbox{c.c.} \Big),\qquad
\end{eqnarray}

\noindent
where $e_0$ and $\lambda_0^{ijk}$ are the bare gauge and Yukawa coupling constants, respectively. In the notation adopted in this paper the supersymmetric covariant derivatives are defined as

\begin{equation}\label{Covariant_Derivatives_Definition}
\nabla_a = \bm{\nabla}_a = D_a;\qquad \bar\nabla_{\dot a} = e^{2{\cal F}(V)} e^{2\bm{V}} \bar D_{\dot a} e^{-2\bm{V}} e^{-2{\cal F}(V)};\qquad  \bm{\bar\nabla}_{\dot a} = e^{2\bm{V}} \bar D_{\dot a} e^{-2\bm{V}}.
\end{equation}

\noindent
They are present inside the regulator functions $R(x)$ and $F(x)$ which should rapidly grow at infinity and satisfy the condition $R(0)=F(0)=1$. The simplest choice is $R(x)=1+x^m$, $F(x)=1+x^n$, where $m$ and $n$ are positive integers. In Eq. (\ref{Action_With_HD_Regularization}) the gauge superfield strength is

\begin{equation}\label{W_Definition_Of}
W_a \equiv \frac{1}{8} \bar D^2 \left(e^{-2\bm{V}} e^{-2{\cal F}(V)}\, D_a \left(e^{2{\cal F}(V)}e^{2\bm{V}}\right)\right).
\end{equation}

As a gauge fixing action we use the term

\begin{equation}\label{Gauge_Fixing_Action}
S_{\mbox{\scriptsize gf}} = -\frac{1}{16\xi_0 e_0^2}\, \mbox{tr} \int d^4x\,d^4\theta\,
\bm{\nabla}^2 V K\Big(-\frac{\bm{\bar\nabla}^2 \bm{\nabla}^2}{16\Lambda^2}\Big)_{Adj} \bm{\bar\nabla}^2 V,
\end{equation}

\noindent
where $K(0)=1$ and $K(x)\to \infty$ at $x\to\infty$. The corresponding Faddeev--Popov and Nielsen--Kallosh actions are written as

\begin{eqnarray}\label{Action_FP_Ghosts}
&& S_{\mbox{\scriptsize FP}} = \frac{1}{2} \int d^4x\, d^4\theta\,
\frac{\partial {\cal F}^{-1}(\widetilde V)^A}{\partial {\widetilde V}^B}\left.\vphantom{\frac{1}{2}}\right|_{\widetilde V = {\cal F}(V)} \left(\big(e^{2\bm{V}}\big)_{Adj}\bar c +
\bar c^+ \right)^A\nonumber\\
&&\qquad\qquad\qquad\qquad \times \left\{\vphantom{\frac{1}{2}}\smash{\Big(\frac{{\cal F}(V)}{1-e^{2{\cal F}(V)}}\Big)_{Adj} c^+
+ \Big(\frac{{\cal F}(V)}{1-e^{-2{\cal F}(V)}}\Big)_{Adj}
\Big(\big(e^{2\bm{V}}\big)_{Adj} c \Big)}\right\}^B;\qquad\\
&& S_{\mbox{\scriptsize NK}} = \frac{1}{2e_0^2}\, \mbox{tr} \int d^4x\,d^4\theta\,  b^+ \Big(K\Big(-\frac{\bm{\bar\nabla}^2 \bm{\nabla}^2}{16\Lambda^2}\Big) e^{2\bm{V}}\Big)_{Adj} b,
\end{eqnarray}

\noindent
respectively. Here $c=e_0 c^A t^A$ and $\bar c$ are the chiral Faddeev--Popov ghost superfields, and the chiral superfield $b=e_0 b^A t^A$ stands for the Nielsen--Kallosh ghosts. Note that after the gauge fixing procedure the total action $S_{\mbox{\scriptsize total}} = S_{\mbox{\scriptsize reg}} + S_{\mbox{\scriptsize gf}} + S_{\mbox{\scriptsize FP}} + S_{\mbox{\scriptsize NK}}$ remains invariant under the background gauge transformations and the BRST transformations \cite{Becchi:1974md,Tyutin:1975qk} in the superfield version \cite{Piguet:1981fb}.

The one-loop divergences are regularized by inserting into the generating functional the relevant Pauli--Villars determinants \cite{Slavnov:1977zf}. In the supersymmetric case, following Ref. \cite{Kazantsev:2017fdc}, the generating functional for the regularized theory can be constructed as

\begin{equation}\label{Z_Generating_Functional}
Z = \int D\mu\, \mbox{Det}(PV,M_\varphi)^{-1} \mbox{Det}(PV,M)^c \exp\Big\{i\Big(S_{\mbox{\scriptsize reg}} + S_{\mbox{\scriptsize gf}} + S_{\mbox{\scriptsize FP}} + S_{\mbox{\scriptsize NK}} + S_{\mbox{\scriptsize sources}}\Big)\Big\},
\end{equation}

\noindent
where $D\mu$ is the measure of the functional integration and

\begin{equation}
\mbox{Det}(PV,M_\varphi)^{-1} \equiv \int D\varphi_1\, D\varphi_2\, D\varphi_3\, \exp(iS_\varphi); \qquad \mbox{Det}(PV,M)^{-1} \equiv \int D\Phi\, \exp(iS_\Phi).
\end{equation}

\noindent
In the first determinant $\varphi_a$ with $a=1,2,3$ are the commuting chiral superfields in the adjoint representation of the gauge group, which cancel one-loop divergences introduced by gauge and ghost loops. The second determinant contains commuting chiral superfields $\Phi_i$ in the representation $R_{\mbox{\scriptsize PV}}$ which admits the gauge invariant mass term with $M^{jk} M^*_{ki} = M^2 \delta_i^j$. (For example, it is possible to choose the adjoint representation.) These superfields cancel the one-loop divergences coming from the loop of the matter superfields $\phi_i$ for $c = T(R)/T(R_{\mbox{\scriptsize PV}})$, where $T(R)$ is defined by the equation $\mbox{tr}(T^A T^B) = T(R) \delta^{AB}$. To regularize the one-loop supergraphs, the actions for the Pauli--Villars superfields should be written as

\begin{eqnarray}\label{S_PV_Varphi}
&&\hspace*{-11mm} S_\varphi = \frac{1}{4} \int d^4x\,d^4\theta\,\Bigg\{ \varphi_1^{*A} \Big[ \Big(R\Big(-\frac{\bar\nabla^2 \nabla^2}{16\Lambda^2}\Big) e^{2{\cal F}(V)} e^{2\bm{V}}\Big)_{Adj} \varphi_1\Big]_A
+ \varphi_2^{*A} \Big[\Big( e^{2{\cal F}(V)} e^{2\bm{V}}\Big)_{Adj} \varphi_2\Big]_A \nonumber\\
&&\hspace*{-11mm} + \varphi_3^{*A} \Big[\Big(e^{2{\cal F}(V)} e^{2\bm{V}}\Big)_{Adj} \varphi_3\Big]_A
\Bigg\} + \Big(\frac{1}{4} M_\varphi \int d^4x\,d^2\theta\, \Big((\varphi_1^A)^2 + (\varphi_2^A)^2 + (\varphi_3^A)^2\Big)+\mbox{c.c}\Big);\\
\label{S_PV_Phi}
&&\hspace*{-11mm} S_\Phi = \frac{1}{4} \int d^4x\,d^4\theta\, \Phi^{*i} \Big(F\Big(-\frac{\bar\nabla^2 \nabla^2}{16\Lambda^2}\Big) e^{2{\cal F}(V)}e^{2\bm{V}}\Big)_i{}^j \Phi_j
+ \Big(\frac{1}{4} M^{ij} \int d^4x\,d^2\theta\, \Phi_i \Phi_j + \mbox{c.c.}\Big).\vphantom{\Bigg(}
\end{eqnarray}

\noindent
The masses of the Pauli--Villars superfields should be proportional to the parameter $\Lambda$ for obtaining a regularized theory with a single dimensionful parameter,

\begin{equation}\label{Coefficients_A}
M_\varphi = a_\varphi \Lambda;\qquad M = a\Lambda,
\end{equation}

\noindent
where the coefficients $a_\varphi$ and $a$ are independent of couplings.

\section{Features of quantum corrections in supersymmetric theories regularized by higher derivatives}
\hspace*{\parindent}\label{Section_Features_Of_Quantum_Corrections}

Application of the higher covariant derivative regularization to explicit calculations revealed a lot of interesting features of quantum corrections in supersymmetric theories which could not be noted with the dimensional reduction. In particular, the calculation of Ref. \cite{Soloshenko:2003nc} demonstrated that the integrals giving the $\beta$-function of ${\cal N}=1$ supersymmetric electrodynamics (SQED) are integrals of total derivatives with respect to the loop momenta. In Ref. \cite{Smilga:2004zr} it was noted that they are also integrals of double total derivatives. The all-order proof of this fact has been done in Refs. \cite{Stepanyantz:2011jy,Stepanyantz:2014ima} and subsequently verified by an explicit three-loop calculation of Ref. \cite{Kazantsev:2014yna}. These integrals are similar to the toy integral

\begin{equation}\label{Singular_Integrals}
\int \frac{d^4Q}{(2\pi)^4} \frac{\partial^2}{\partial Q_\mu^2} \Big[\frac{f(Q^2)}{Q^2}\Big] = \frac{1}{8\pi^4} \oint\limits_{S^3_\varepsilon} dS\, \frac{1}{Q^3}\, \Big(f(Q^2) - Q^2 f'(Q^2)\Big) = \frac{1}{4\pi^2} f(0),
\end{equation}

\noindent
where $f(Q^2)$ is a nonsingular function rapidly decreasing at infinity. Due to the factorization into double total derivatives one of loop integrals can be calculated analytically. In the Abelian case this allows to relate the $\beta$-function in a certain order to the anomalous dimension of the matter superfields in the previous order by the so-called exact NSVZ $\beta$-function \cite{Vainshtein:1986ja,Shifman:1985fi}

\begin{equation}\label{NSVZ_Abelian}
\beta(\alpha_0) \equiv \left.\frac{d\alpha_0}{d\ln\Lambda}\right|_{\alpha=\mbox{\scriptsize const}} = \frac{\alpha_0^2 N_f}{\pi}\Big(1-\gamma(\alpha_0)\Big) \equiv \frac{\alpha_0^2 N_f}{\pi}\bigg(1+\left.\frac{d\ln Z}{d\ln\Lambda}\right|_{\alpha=\mbox{\scriptsize const}} \bigg),
\end{equation}

\noindent
where $N_f$ is a number of flavors \cite{Stepanyantz:2011jy,Stepanyantz:2014ima}. It is important that the renormalization group functions (RGFs) in this equation are defined in terms of the bare coupling constant. Such RGFs are scheme independent for a fixed regularization, so that in the case of using the higher derivative regularization Eq. (\ref{NSVZ_Abelian}) takes place in all orders for an arbitrary renormalization prescription. In the case of using the dimensional reduction this is not so \cite{Aleshin:2015qqc,Aleshin:2016rrr}, and Eq. (\ref{NSVZ_Abelian}) for RGFs defined in terms of the bare coupling constant does not take place starting from the three-loop approximation.

For the standard definition of RGFs in terms of the renormalized coupling constant,

\begin{equation}
\widetilde\beta(\alpha) \equiv \left.\frac{d\alpha}{d\ln\mu}\right|_{\alpha_0=\mbox{\scriptsize const}}; \qquad \widetilde\gamma(\alpha) \equiv \left.\frac{d\ln Z}{d\ln\mu}\right|_{\alpha_0=\mbox{\scriptsize const}},
\end{equation}

\noindent
the Abelian NSVZ equation is valid only in a certain class of the subtraction schemes described in \cite{Goriachuk:2018cac}. It is well known that the $\overline{\mbox{DR}}$-scheme (i.e. dimensional reduction supplemented by the modified minimal subtractions \cite{Bardeen:1978yd}) does not belong to this class in both Abelian and non-Abelian cases \cite{Jack:1996vg,Jack:1996cn,Jack:1998uj,Harlander:2006xq,Mihaila:2013wma}.

However, the NSVZ scheme can easily be constructed in all orders in the case of using the higher derivative regularization \cite{Kataev:2013eta,Kataev:2013csa,Kataev:2014gxa}. It is obtained with the so-called HD+MSL prescription \cite{Shakhmanov:2017wji,Stepanyantz:2017sqg}, when the theory is regularized by Higher covariant Derivatives and Minimal Subtractions of Logarithms are used for the renormalization. In this case one should include into the renormalization constants only powers of $\ln\Lambda/\mu$. Another example of an all-order prescription which gives the NSVZ scheme in the Abelian case is the on-shell scheme \cite{Kataev:2019olb}. The MOM scheme is not NSVZ \cite{Kataev:2014gxa}.

Using the method proposed in Ref. \cite{Stepanyantz:2011jy} some NSVZ-like relations were also proved in all orders. In particular, the factorization of integrands into double total derivatives turns out to produce the
NSVZ-like equation \cite{Hisano:1997ua,Jack:1997pa,Avdeev:1997vx} describing running of the photino mass in softly broken SQED \cite{Nartsev:2016nym} and the NSVZ-like relation for the Adler $D$-function in ${\cal N}=1$ SQCD \cite{Shifman:2014cya,Shifman:2015doa}. In both cases with the higher covariant derivative regularization RGFs defined in terms of the bare couplings satisfy these equations independently of a renormalization prescripton, while the NSVZ scheme for RGFs defined in terms of the renormalized couplings is given by the HD+MSL prescription, see Refs. \cite{Nartsev:2016mvn} and \cite{Kataev:2017qvk}. In the case of using dimensional reduction the former RGFs do not satisfy the NSVZ equation, while for the latter RGFs the NSVZ scheme should be specially tuned in each order of the perturbation theory, see Refs. \cite{Jack:1997pa} and \cite{Aleshin:2019yqj}.

However, the generalization of the results obtained in Ref. \cite{Stepanyantz:2011jy} to the non-Abelian case turned out to be a much more complicated problem. Although numerous explicit calculations made with the higher covariant derivative regularization \cite{Pimenov:2009hv,Stepanyantz:2011cpt,Stepanyantz:2011zz,Stepanyantz:2011bz,Aleshin:2016yvj,Shakhmanov:2017soc,Kazantsev:2018nbl} reveal the structure of loop integrals similar to the Abelian case, the factorization into the double total derivatives has been proved in all orders only recently \cite{Stepanyantz:2019ihw}. (Note that the main ingredient of this proof is again the regularization by higher covariant derivatives.) Moreover, the non-Abelian NSVZ equation \cite{Novikov:1983uc,Jones:1983ip,Novikov:1985rd,Shifman:1986zi}\footnote{In Eqs. (\ref{NSVZ_Equation_Original_Form}) and (\ref{NSVZ_Equation_New_Form}) we do not specify the definitions of RGFs.}

\begin{equation}\label{NSVZ_Equation_Original_Form}
\beta(\alpha,\lambda) = - \frac{\alpha^2\Big(3 C_2 - T(R) + C(R)_i{}^j \big(\gamma_\phi\big)_j{}^i(\alpha,\lambda)/r\Big)}{2\pi(1- C_2\alpha/2\pi)},
\end{equation}

\noindent
where $r\equiv \mbox{dim}\, G$,\ \ $C(R)_i{}^j\equiv (T^A T^A)_i{}^j$, and $C_2\equiv T(Adj)$, relates the $\beta$-function to the anomalous dimension of matter superfields in all previous orders, while calculating integrals of double total derivatives we reduce the number of loop integrals only by 1. The solution of this problem has been found using the new non-renormalization theorem for the triple gauge-ghost vertices proved in Ref. \cite{Stepanyantz:2016gtk} using the Slavnov--Taylor identities \cite{Taylor:1971ff,Slavnov:1972fg} and the Feynman rules in ${\cal N}=1$ superspace \cite{Gates:1983nr,West:1990tg,Buchbinder:1998qv}.\footnote{Earlier similar statements were known only in the Landau gauge $\xi\to 0$ for the (non-supersymmetric) Yang--Mills theory \cite{Dudal:2002pq} and for ${\cal N}=1$ SYM in the Wess-Zumiino gauge \cite{Capri:2014jqa}, while the superfield results of Ref. \cite{Stepanyantz:2016gtk} are valid for the general $\xi$-gauge.} This non-renormalization theorem states that the vertices with two external ghost legs and one external leg of the quantum gauge superfield are finite in all orders of the perturbation theory. This result can be written in the form

\begin{equation}
\frac{d}{d\ln\Lambda} (Z_\alpha^{-1/2} Z_c Z_V) = 0,
\end{equation}

\noindent
where the renormalization constants are defined by the equations

\begin{equation}\label{Definition_Of_Zs}
\frac{1}{\alpha_0} = \frac{Z_\alpha}{\alpha};\qquad V = Z_V Z_\alpha^{-1/2} V_R;\qquad \bar c c = Z_c Z_\alpha^{-1} \bar c_R c_R;\qquad \phi_i = (\sqrt{Z_\phi})_i{}^j (\phi_R)_j.
\end{equation}

\noindent
Therefore, there is a subtraction scheme in which $Z_\alpha^{-1/2} Z_c Z_V=1$. This allows rewriting the exact NSVZ $\beta$-function in a new form, which expresses it in terms of the anomalous dimensions of the quantum superfields \cite{Stepanyantz:2016gtk},

\begin{equation}\label{NSVZ_Equation_New_Form}
\frac{\beta(\alpha,\lambda)}{\alpha^2} = - \frac{1}{2\pi}\Big(3 C_2 - T(R) - 2C_2 \gamma_c(\alpha,\lambda) - 2C_2 \gamma_V(\alpha,\lambda) + C(R)_i{}^j \big(\gamma_\phi\big)_j{}^i(\alpha,\lambda)/r\Big).
\end{equation}

\noindent
It is important that this equation relates the $\beta$-function in a certain order only to the anomalous dimensions in the previous order and has the same qualitative interpretation as in the Abelian case. Namely, if we consider a supergraph without external legs, then attaching two external $\bm{V}$-lines in all possible ways we obtain a (rather large) set of two-point superdiagrams contributing to the function $\beta(\alpha_0,\lambda_0)/\alpha_0^2$. From the other side, by cutting internal lines in the original graph we produce a set of superdiagrams contributing to the anomalous dimensions of the quantum gauge superfield $\gamma_V(\alpha_0,\lambda_0)$, of the Faddeev--Popov ghost $\gamma_c(\alpha_0,\lambda_0)$, and of the matter superfields $(\gamma_\phi)_i{}^j(\alpha_0,\lambda_0)$. Then, all these contributions to various RGFs are related by Eq. (\ref{NSVZ_Equation_New_Form}).

Due to the factorization into double total derivatives Eq. (\ref{NSVZ_Equation_New_Form}) presumably takes place for RGFs defined in terms of the bare couplings in all loops in the case of using the higher covariant derivative regularization. Then according to \cite{Stepanyantz:2016gtk}, the NSVZ scheme in the non-Abelian case is given by the HD+MSL prescription. Although the general all-loop proof of these facts is in preparation, the explicit calculations of Refs. \cite{Shakhmanov:2017soc,Kazantsev:2018nbl} confirm them. In these papers the three-loop contributions to the $\beta$-function containing the Yukawa couplings have been compared with the corresponding two-loop contributions to the anomalous dimensions of the quantum superfields. The NSVZ equations (\ref{NSVZ_Equation_Original_Form}) and (\ref{NSVZ_Equation_New_Form}) have been checked for RGFs defined in terms of the bare couplings and for RGFs defined in terms of the renormalized couplings in the HD+MSL scheme. It is important that in this approximation the scheme dependence becomes essential.\footnote{For RGFs defined in terms of the bare couplings the scheme dependence is reduced to the regularization dependence.} Certainly, the corresponding calculations are very complicated from the technical point of view. That is why the complete three-loop calculation with the higher covariant derivative has not yet been done. Actually, only the one-loop calculation with the regularization described above has been made in Ref. \cite{Aleshin:2016yvj}. However, according to Ref. \cite{Stepanyantz:2019ihw} it is possible to construct a special method for calculating the $\beta$-function with the higher covariant derivative regularization which simplifies the calculations in a great extent and produces the result in the form of integrals of double total derivatives. Below we formulate the corresponding algorithm and apply it for calculating the two-loop $\beta$-function.

\section{An algorithm for simple calculating the $\beta$-function with the higher covariant derivative regularization}
\hspace*{\parindent}\label{Section_Algorithm}

The proof of the factorization of the loop integrals giving the $\beta$-function into integrals of double total derivatives suggests a simple method for constructing these integrals \cite{Stepanyantz:2019ihw,Kuzmichev:2019ywn}. It is based on the observation that integrals of double total derivatives appear even for a sum of superdiagrams which are formally obtained from a single vacuum supergraph by attaching two external lines of the background gauge superfield in all possible ways. So, let us start with a certain $L$-loop supergraph without external lines and describe an algorithm for obtaining the corresponding contribution to the function $\beta(\alpha_0,\lambda_0)/\alpha_0^2$:

1. With the help of supersymmetric Feynman rules we should formally write the contribution to the effective action coming from the supergraph under consideration.

2. After this, it is necessary to insert the expression $\theta^4 (v^B)^2$ to an arbitrary point containing the integration over $d^4\theta$, where $v^B$ are the slowly changing functions which tend to 0 only at a very large scale $R$. If there are no such points, then one should convert one of the integrations over $d^2\theta$ into $d^4\theta$, which is always possible.

3. Then one should calculate the obtained expression omitting terms suppressed by powers of $1/(\Lambda R)$ and extract the factor

\begin{equation}
{\cal V}_4 \equiv \int d^4x\, \big(v^B\big)^2 \to \infty.
\end{equation}

4. Next, we mark $L$ propagators corresponding to the Euclidean momenta $Q_i^\mu$ which are considered as independent. Let $a_i$ be the indices corresponding to their beginnings. Evidently, the product of these propagators is proportional to $\prod_{i=1}^L \delta_{a_i}^{b_i}$.

5. The loop integral corresponding to the considered vacuum supergraph is modified by the formal replacement

\begin{equation}
\prod\limits_{i=1}^L \delta_{a_i}^{b_i} \ \to\ \sum_{k,l=1}^L  \prod\limits_{i\ne k,l} \delta_{a_i}^{b_i}\, (T^A)_{a_k}{}^{b_k} (T^A)_{a_l}{}^{b_l} \frac{\partial^2}{\partial Q^\mu_k \partial Q^\mu_l}
\end{equation}

\noindent
in the integrand.

6. Finally the result is multiplied by

\begin{equation}
- \frac{2\pi}{r{\cal V}_4} \frac{d}{d\ln\Lambda}.
\end{equation}

\noindent
Note that the differentiation with respect to $\ln\Lambda$ should be made before the integration over loop momenta to avoid appearing of expressions which diverge in the infrared region.

As a result of the procedure described above we obtain a part of the expression

\begin{equation}
\frac{1}{\alpha_0^2}\Big(\beta(\alpha_0,\lambda_0) -\beta_{\mbox{\scriptsize 1-loop}}(\alpha_0) \Big)
\end{equation}

\noindent
coming from all two-point superdiagrams corresponding to the original vacuum supergraph.

This algorithm simplifies the calculations to a great extent, because instead of calculating a large number of two-point superdiagrams one has to calculate only a single supergraph without external legs.

\section{Two-loop $\beta$-function with the higher derivative regularization}
\hspace*{\parindent}\label{Section_Two_Loop_Beta_Function}

As an illustration of the method described above, we calculate the two-loop $\beta$-function of the theory (\ref{Superfield_Action}) in the general $\xi$-gauge (\ref{Gauge_Fixing_Action}). It is worth to note that earlier the two-loop $\beta$-function has been obtained only in the Feynman gauge $\xi=1$ in the case of using the simplified version of the higher derivative regularization breaking the BRST invariance. To restore the Slavnov--Taylor identities \cite{Taylor:1971ff,Slavnov:1972fg} in this case one had to use a special renormalization procedure proposed in \cite{Slavnov:2001pu,Slavnov:2002ir} and subsequently generalized to the supersymmetric case in \cite{Slavnov:2002kg,Slavnov:2003cx}. The result found in Ref. \cite{Pimenov:2009hv}\footnote{Similar calculations with different non-invariant versions of the higher derivative regularization can be found in Refs. \cite{Stepanyantz:2011cpt,Stepanyantz:2011zz}.} using the standard technique of calculations was originally written as a sum of integrals of total derivatives. Subsequently, it was noted that it can be also written in the form of integrals of double total derivatives \cite{Stepanyantz:2011bz}. Comparing the result with Eq. (\ref{NSVZ_Equation_New_Form}), it is possible to see \cite{Shakhmanov:2017wji} that Eq. (\ref{NSVZ_Equation_New_Form}) really relates the two-loop $\beta$-function to the one-loop anomalous dimensions of the quantum superfields at the level of loop integrals.

However, the two-loop $\beta$-function has never been calculated with the (BRST invariant) version of the higher covariant derivative regularization described in this paper. Moreover, the gauge dependence of NSVZ equation has not also been investigated. In the case of using the standard methods, the corresponding calculation appears to be rather complicated. However, the technique described above allows making it much easier.

\begin{figure}[h]
\begin{picture}(0,5.9)
\put(1.3,3.2){\includegraphics[scale=0.2]{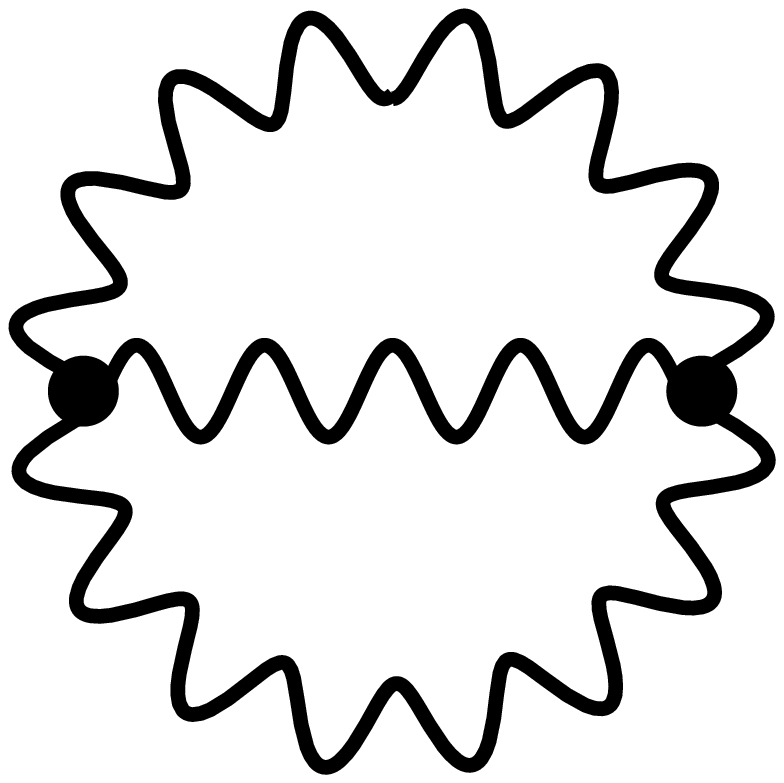}}
\put(4.3,3.2){\includegraphics[scale=0.225]{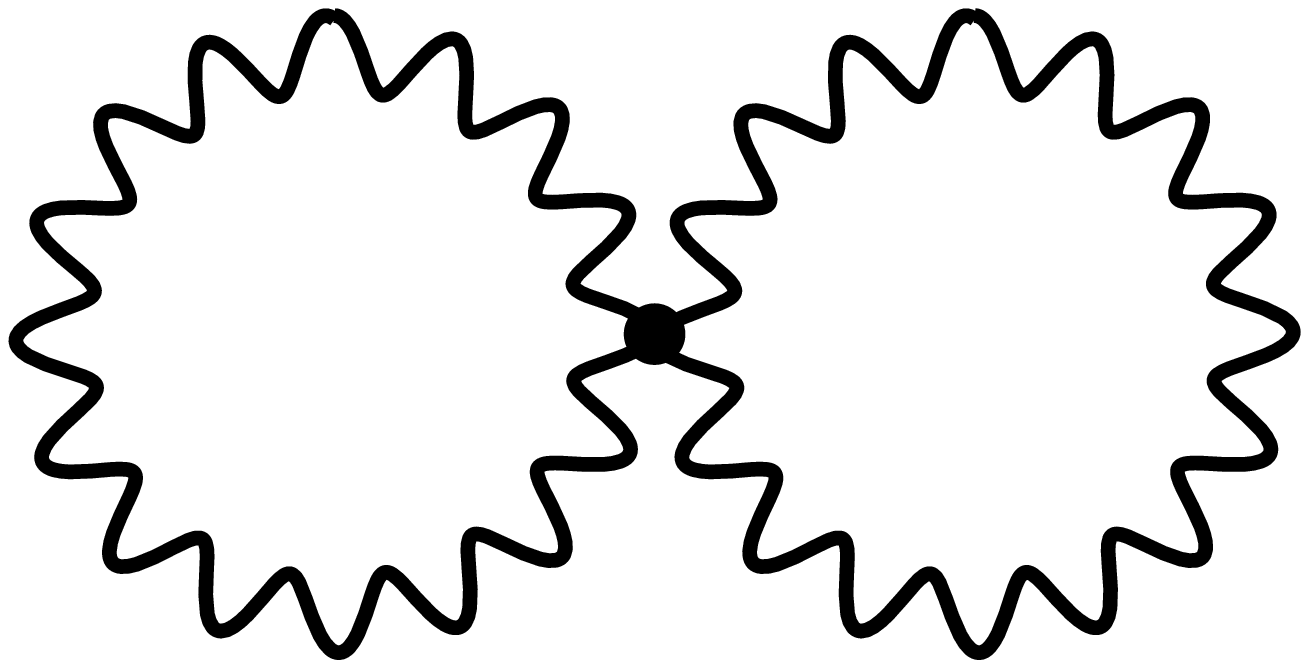}}
\put(8.9,3.2){\includegraphics[scale=0.225]{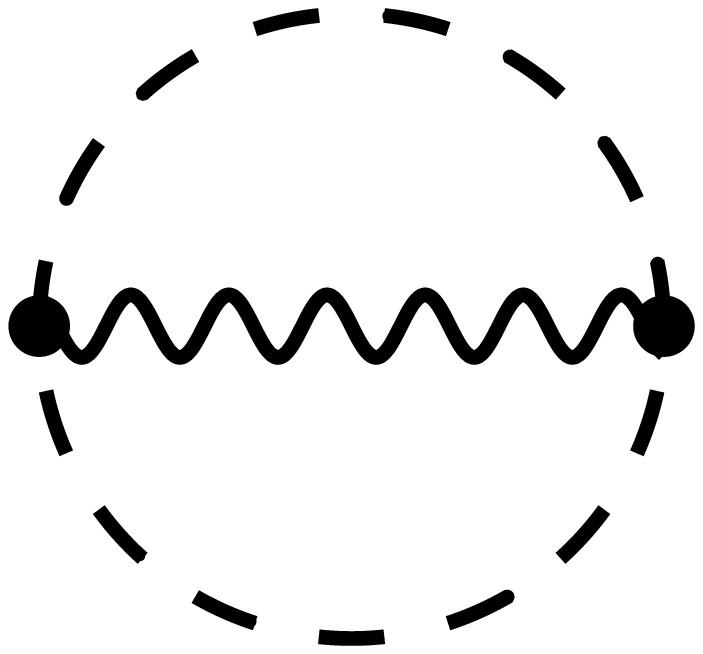}}
\put(12.0,3.2){\includegraphics[scale=0.22]{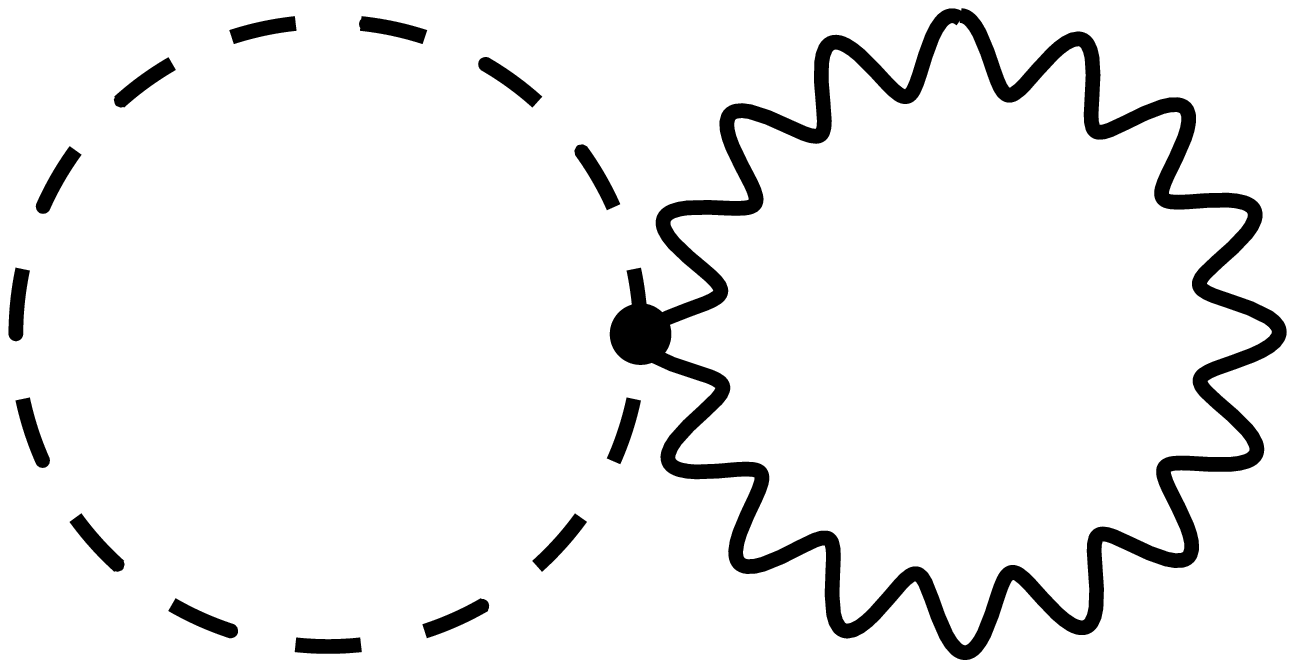}}
\put(3.8,0.2){\includegraphics[scale=0.185]{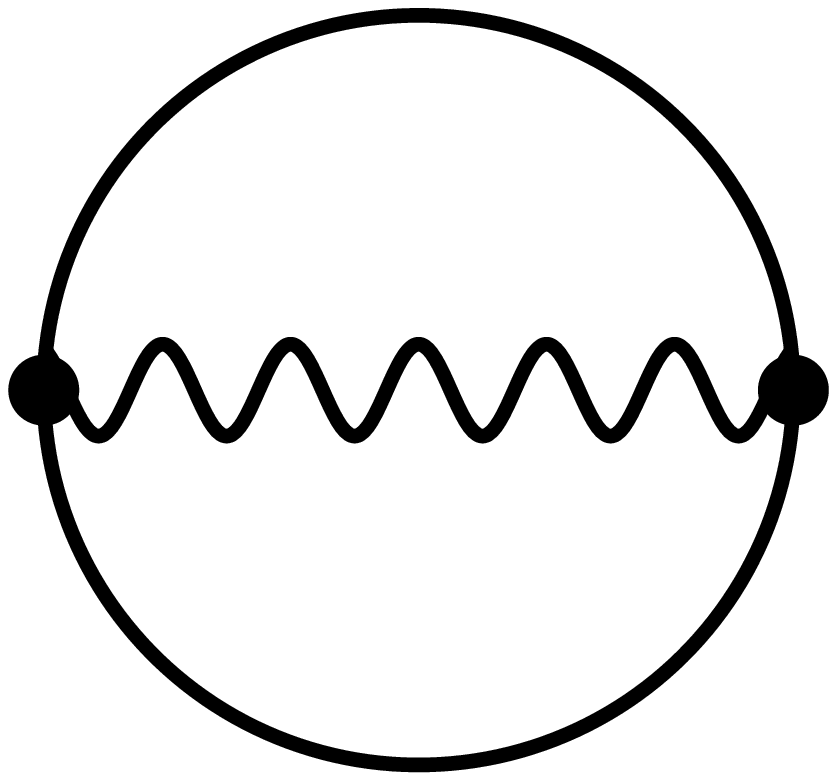}}
\put(6.93,0.2){\includegraphics[scale=0.218]{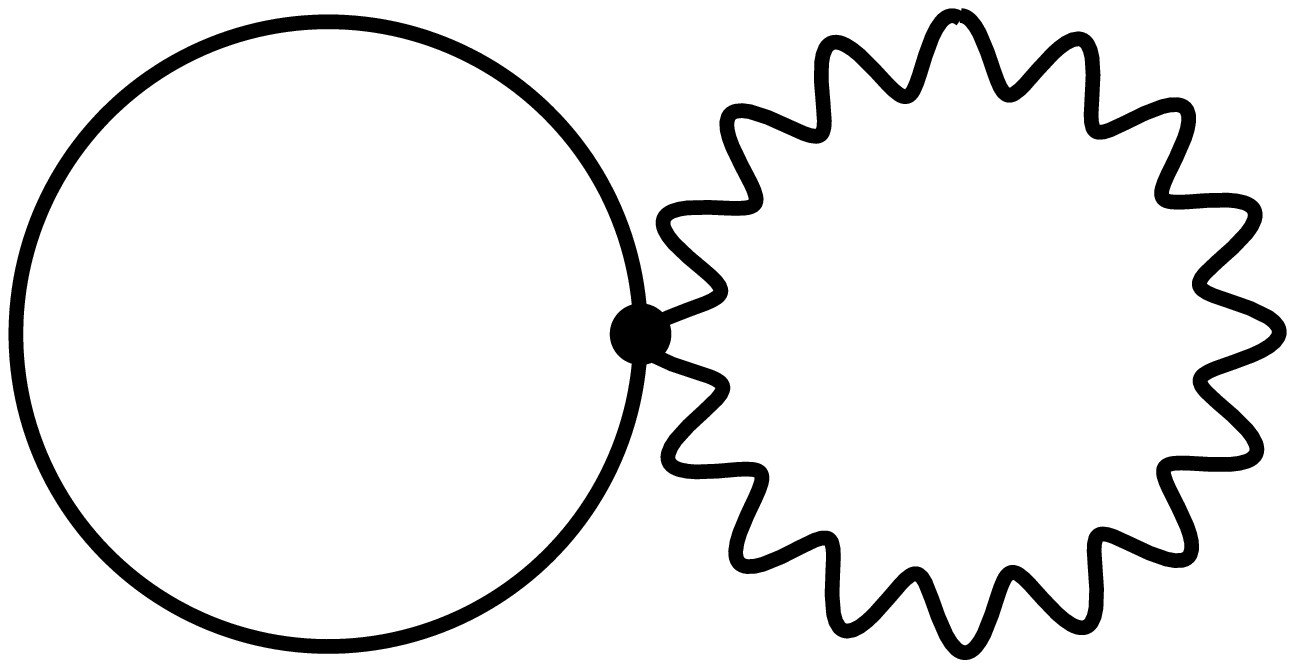}}
\put(11.3,0.2){\includegraphics[scale=0.218]{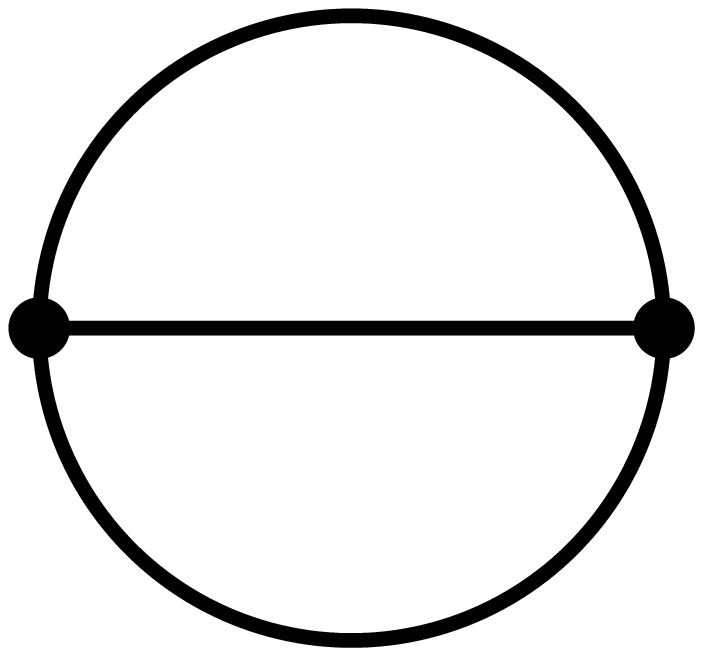}}
\put(1,4.9){B1} \put(4.1,4.9){B2} \put(8.6,4.9){B3} \put(11.7,4.9){B4}
\put(3.45,1.8){B5} \put(6.6,1.8){B6} \put(11,1.8){B7}
\end{picture}
\caption{Supergraphs generating the two-loop contributions to the $\beta$-function. The corresponding two-point superdiagrams are obtained from them by attaching two external $\bm{V}$-legs in all possible ways.}\label{Figure_Beta_Supergraphs}
\end{figure}

The two-loop contribution to the $\beta$-function is generated by the supergraphs presented in Fig.~\ref{Figure_Beta_Supergraphs}. To obtain (a very large number of) usual two-point superdiagrams, one should attach to them two external legs of the background gauge superfield in all possible ways. However, as we discussed earlier, in the case of using the higher covariant derivative regularization their contributions can be constructed by calculating only specially modified vacuum supergraphs presented in Fig. \ref{Figure_Beta_Supergraphs}. The result of this calculation obtained according to the algorithm described above can be written as

\begin{eqnarray}
&&\hspace*{-5mm} \frac{\beta(\alpha_0,\lambda_0)}{\alpha_0^2} - \frac{\beta_{\mbox{\scriptsize 1-loop}}(\alpha_0)}{\alpha_0^2} = \Delta_{\mbox{\scriptsize gauge+ghost}}\Big(\frac{\beta}{\alpha_0^2}\Big) + \Delta_{\varphi}\Big(\frac{\beta}{\alpha_0^2}\Big) + \Delta_{\mbox{\scriptsize matter}}\Big(\frac{\beta}{\alpha_0^2}\Big) + \Delta_{\mbox{\scriptsize Yukawa}}\Big(\frac{\beta}{\alpha_0^2}\Big)\nonumber\\
&&\hspace*{-5mm} + O\Big(\alpha_0^2,\alpha_0\lambda_0^2,\lambda_0^4\Big),
\end{eqnarray}

\noindent
where the expressions for all $\Delta_I(\beta/\alpha_0^2)$ are presented in Appendix \ref{Appendix_Results_For_Supergraphs} and

\begin{equation}\label{Beta_In_The_One_Loop}
\beta_{\mbox{\scriptsize 1-loop}}(\alpha_0) = - \frac{\alpha_0^2}{2\pi}\Big(3C_2-T(R)\Big)
\end{equation}

\noindent
is the one-loop contribution to the $\beta$-function. With the regularization considered in this paper it has been calculated in Ref. \cite{Aleshin:2016yvj}. The contributions $\Delta_{\mbox{\scriptsize gauge+ghost}}(\beta/\alpha_0^2)$, $\Delta_{\mbox{\scriptsize matter}}(\beta/\alpha_0^2)$, and $\Delta_{\mbox{\scriptsize Yukawa}}(\beta/\alpha_0^2)$ are the sums of the superdiagrams generated by the supergraphs B1 -- B4, B5 and B6 (including the supergraphs containing a loop of the Pauli--Villars superfields $\Phi_i$), and B7, respectively. The contribution $\Delta_{\varphi}(\beta/\alpha_0^2)$ comes from the supergraphs B5 and B6 in which the solid line corresponds to a loop of the Pauli--Villars superfields $\varphi_a$.

It is necessary to note that all gauge dependent terms containing the parameter $\xi_0$ and the regulator $K(x)$ (see Eq. (\ref{Gauge_Fixing_Action})) cancel each other, so that the resulting expression turns out to be gauge independent. However, the expressions for separate supergraphs contain gauge dependent terms. (For example, the results for the supergraphs B3 and B4 explicitly written in Ref. \cite{Kuzmichev:2019ywn} for the particular case $K(x)=R(x)$ do contain them.) Also it should be noted that the expressions for separate supergraphs contain some terms which are not well defined. However, in their sum all bad terms disappear and the resulting expression is well defined.

With the help of equations similar to (\ref{Singular_Integrals}) it is possible to calculate one of the loop integrals in the expressions (\ref{Contribution_Gauge+Ghost}) --- (\ref{Contribution_Yukawa}). Omitting inessential higher order terms we obtain

\begin{eqnarray}
&& \Delta_{\mbox{\scriptsize gauge+ghost}}\Big(\frac{\beta}{\alpha_0^2}\Big) =  \alpha_0 C_2^2 \int \frac{d^4Q}{(2\pi)^4} \frac{d}{d\ln\Lambda} \frac{\partial^2}{\partial Q_\mu^2}\Big[\frac{1}{Q^2}\ln R_Q\Big];\\
&& \Delta_\varphi\Big(\frac{\beta}{\alpha_0^2}\Big) = -\alpha_0 C_2^2  \int \frac{d^4Q}{(2\pi)^4} \frac{d}{d\ln\Lambda} \frac{\partial^2}{\partial Q_\mu^2} \left[\frac{1}{Q^2}\smash{\ln\Big(1+\frac{M_\varphi^2}{Q^2}\Big)}\right.\nonumber\\
&&\qquad\qquad\qquad\qquad\qquad\qquad\qquad\qquad\qquad\quad\  \left. + \frac{1}{2Q^2} \smash{\ln\Big(1+\frac{M_\varphi^2}{Q^2 R_Q^2}\Big)} + \frac{1}{Q^2} \ln R_Q\right];\\
&& \Delta_{\mbox{\scriptsize matter}}\Big(\frac{\beta}{\alpha_0^2}\Big) = \frac{4\alpha_0}{r}\, \mbox{tr}\left(C(R)^2\right) \int\frac{d^4Q}{(2\pi)^4} \frac{d}{d\ln\Lambda} \frac{1}{Q^4 R_Q}\nonumber\\
&&\qquad\qquad\qquad\qquad\qquad + \frac{\alpha_0}{2r}\, C_2\, \mbox{tr}\, C(R) \int \frac{d^4Q}{(2\pi)^4} \frac{d}{d\ln\Lambda} \frac{\partial^2}{\partial Q_\mu^2} \left[\frac{1}{Q^2}\smash{\ln\Big(1+\frac{M^2}{Q^2 F_Q^2}\Big)}\right];\qquad\\
&& \Delta_{\mbox{\scriptsize Yukawa}}\Big(\frac{\beta}{\alpha_0^2}\Big) = -\frac{1}{\pi r} \lambda^*_{0jmn} \lambda_0^{imn} C(R)_i{}^j  \int \frac{d^4Q}{(2\pi)^4} \frac{d}{d\ln\Lambda} \frac{1}{Q^4 F_Q^2}.
\end{eqnarray}

\noindent
Summing up these contributions and substituting the expression (\ref{Beta_In_The_One_Loop}) for the one-loop $\beta$-function, we find the complete expression for the two-loop $\beta$-function in the form of the loop integrals

\begin{eqnarray}
&& \frac{\beta(\alpha_0,\lambda_0)}{\alpha_0^2} = - \frac{1}{2\pi}\Big(3C_2-T(R)\Big)
-\alpha_0 C_2^2 \int \frac{d^4Q}{(2\pi)^4} \frac{d}{d\ln\Lambda} \frac{\partial^2}{\partial Q_\mu^2} \left[\frac{1}{2Q^2} \smash{\ln\Big(1+\frac{M_\varphi^2}{Q^2 R_Q^2}\Big)}\right.\nonumber\\
&& \left. +\frac{1}{Q^2}\smash{\ln\Big(1+\frac{M_\varphi^2}{Q^2}\Big)}\right] + \frac{\alpha_0}{2r}\, C_2\, \mbox{tr}\, C(R) \int \frac{d^4Q}{(2\pi)^4} \frac{d}{d\ln\Lambda} \frac{\partial^2}{\partial Q_\mu^2} \left[\frac{1}{Q^2}\smash{\ln\Big(1+\frac{M^2}{Q^2 F_Q^2}\Big)}\right]\nonumber\\
&& +\frac{4\alpha_0}{r} \mbox{tr}\left(C(R)^2\right) \int\frac{d^4Q}{(2\pi)^4} \frac{d}{d\ln\Lambda} \frac{1}{Q^4 R_Q} -\frac{1}{\pi r} \lambda^*_{0jmn} \lambda_0^{imn} C(R)_i{}^j \int \frac{d^4Q}{(2\pi)^4} \frac{d}{d\ln\Lambda} \frac{1}{Q^4 F_Q^2}\qquad\nonumber\\
&& + O\big(\alpha_0^2,\alpha_0\lambda_0^2,\lambda_0^4\big).\vphantom{\frac{1}{2}}\qquad
\end{eqnarray}

\noindent
Before calculating the remaining integrals, it is expedient to compare this expression with the one-loop anomalous dimensions of the quantum superfields (defined in terms of the bare couplings) obtained in Ref. \cite{Aleshin:2016yvj},

\begin{eqnarray}\label{Gamma_V_Integral}
&&\hspace*{-3mm} \gamma_V = -\pi\alpha_0 \int\frac{d^4Q}{(2\pi)^4} \frac{d}{d\ln\Lambda} \frac{\partial^2}{\partial Q_\mu^2}\left[\,\frac{C_2}{2Q^2}\,\smash{\ln\Big(1+\frac{M_\varphi^2}{Q^2 R_Q^2}\Big)} + \frac{C_2}{Q^2}\,\smash{\ln\Big(1+\frac{M_\varphi^2}{Q^2}\Big)} - \frac{1}{2r}\mbox{tr}\, C(R)\, \frac{1}{Q^2}\right. \nonumber\\
&&\hspace*{-3mm}\quad\, \left. \times \smash{\ln\Big(1+\frac{M^2}{Q^2 F_Q^2}\Big)}\vphantom{\frac{1}{2}}\right] +4\pi\alpha_0 C_2 \int \frac{d^4Q}{(2\pi)^4} \frac{d}{d\ln\Lambda} \Big(\frac{1}{3 Q^4 R_Q} - \frac{\xi_0}{3Q^4 K_Q}\Big)+ O\big(\alpha_0^2,\alpha_0\lambda_0^2\big);\qquad\\
\label{Gamma_C_Integral}
&&\hspace*{-3mm} \gamma_c = 4\pi\alpha_0 C_2 \int \frac{d^4Q}{(2\pi)^4} \frac{d}{d\ln\Lambda} \Big(-\frac{1}{3 Q^4 R_Q} + \frac{\xi_0}{3Q^4 K_Q}\Big) + O\big(\alpha_0^2,\alpha_0\lambda_0^2\big);\\
\label{Gamma_Phi_Integral}
&&\hspace*{-3mm} \big(\gamma_\phi\big)_i{}^j = \int \frac{d^4Q}{(2\pi)^4} \frac{d}{d\ln\Lambda} \Big(-C(R)_i{}^j \frac{8\pi\alpha_0}{Q^4 R_Q} + \lambda^*_{0imn} \lambda_0^{jmn} \frac{2}{Q^4 F_Q^2}\Big) + O\big(\alpha_0^2,\alpha_0\lambda_0^2,\lambda_0^4\big).\qquad
\end{eqnarray}

\noindent
Then we see that even at the level of the loop integrals the NSVZ equation in the form (\ref{NSVZ_Equation_New_Form}) is satisfied by these RGFs in the considered approximation,

\begin{eqnarray}\label{NSVZ_New_Check}
&&\hspace*{-5mm} \frac{\beta(\alpha_0,\lambda_0)}{\alpha_0^2} = - \frac{1}{2\pi}\Big(3 C_2 - T(R) - 2C_2 \gamma_c(\alpha_0,\lambda_0) - 2C_2 \gamma_V(\alpha_0,\lambda_0) + C(R)_i{}^j \big(\gamma_\phi\big)_j{}^i(\alpha_0,\lambda_0)/r\Big)\nonumber\\
&&\hspace*{-5mm} + O\big(\alpha_0^2,\alpha_0\lambda_0^2,\lambda_0^4\big) = - \frac{1}{2\pi}\Big(3 C_2 - T(R)\Big) + \frac{\alpha_0}{(2\pi)^2}\Big(-3C_2^2 + \frac{1}{r}\, C_2\, \mbox{tr}\, C(R) + \frac{2}{r}\,\mbox{tr}\left(C(R)^2\right)\Big)\nonumber\\
&&\hspace*{-5mm} - \frac{1}{8\pi^3 r} C(R)_i{}^j \lambda^*_{0jmn} \lambda_0^{imn} + O\big(\alpha_0^2,\alpha_0\lambda_0^2,\lambda_0^4\big).
\end{eqnarray}

\noindent
It is easy to verify that (up to notations) this result agrees with the first two-loop calculation made in Ref. \cite{Jones:1974pg} with dimensional regularization. Certainly, this occurs due to the scheme independence of the two-loop $\beta$-function.

Note that, according to \cite{Aleshin:2016yvj}, at the level of loop integrals

\begin{equation}
\gamma_V(\alpha_0,\lambda_0) + \gamma_c(\alpha_0,\lambda_0) = \frac{\beta(\alpha_0,\lambda_0)}{2\alpha_0} + O(\alpha_0^2,\alpha_0\lambda_0^2).
\end{equation}

\noindent
This implies that the original NSVZ equation (\ref{NSVZ_Equation_Original_Form}) is also valid for RGFs defined in terms of the bare couplings in the considered approximation,

\begin{equation}
\frac{\beta(\alpha_0,\lambda_0)}{\alpha_0^2} = - \frac{3 C_2 - T(R) + C(R)_i{}^j \big(\gamma_\phi\big)_j{}^i(\alpha_0,\lambda_0)/r}{2\pi(1- C_2\alpha_0/2\pi)} + O(\alpha_0^2,\alpha_0\lambda_0^2,\lambda_0^4).
\end{equation}

\noindent
This result is analogous to the one obtained in Ref. \cite{Shakhmanov:2017wji} in the case of using the simplified BRST non-invariant version of the higher derivative regularization. However, using of the invariant regularization is certainly more preferable. Moreover, Eq. (\ref{NSVZ_New_Check}) has been derived for the general $\xi$-gauge (\ref{Gauge_Fixing_Action}). This allows investigating the gauge dependence of the NSVZ equation. We see that the $\beta$-function is gauge independent, while the anomalous dimensions (\ref{Gamma_V_Integral}) and (\ref{Gamma_C_Integral}) explicitly depend on the gauge parameter $\xi_0$. However, this gauge dependence disappears in the sum of terms in the right hand side. This implies that the NSVZ equation takes place in an arbitrary $\xi$-gauge.

Taking into account that the two-loop $\beta$-function and the one-loop anomalous dimensions are scheme independent, we also see that in the considered approximation the NSVZ equations (\ref{NSVZ_Equation_Original_Form}) and (\ref{NSVZ_Equation_New_Form}) are also satisfied by RGFs defined in terms of the renormalized couplings for an arbitrary renormalization prescription.

\section*{Conclusion}
\hspace*{\parindent}

In this paper we argue that the Slavnov higher covariant derivative regularization is a very useful tool for both investigating the structure of quantum corrections in supersymmetric theories and explicit perturbative calculations. Unlike the dimensional reduction, it allows to reveal the origin of the exact NSVZ $\beta$-function, namely, the factorization of integrals giving the $\beta$-function into integrals of double total derivatives with respect to the loop momenta. Due to this fact RGFs defined in terms of the bare couplings satisfy the NSVZ equation in all loops independently of a renormalization prescription. For RGFs defined in terms of the renormalized couplings the NSVZ scheme in all orders is given by the HD+MSL prescription, which in particular includes this regularization. Note that the proof of these facts essentially involves the new non-renormalization theorem for the triple gauge-ghost vertices. This theorem was also derived with the help of the higher covariant derivative regularization using the supergraph calculation rules and the Slavnov--Taylor identities.

Last years a large number of explicit multiloop calculations were made with the higher covariant derivative regularization. These calculations confirm the general statements discussed above. In particular, the statement ``HD+MSL=NSVZ'' has been checked in such an order of the perturbation theory where the scheme dependence is essential. Note that recently a special method for making such calculations was proposed \cite{Stepanyantz:2019ihw}. It allows to simplify the calculations considerably and to obtain the results for contributions to $\beta$-function in the form of integrals of double total derivatives. In this paper we demonstrated the application of this method by calculating the two-loop $\beta$-function for the general renormalizable ${\cal N}=1$ supersymmetric gauge theory in the $\xi$-gauge. This calculation confirmed the validity of the new form of the NSVZ equation (\ref{NSVZ_Equation_New_Form}) at the level of loop integrals for an arbitrary $\xi$ and showed the gauge independence of both sides of the NSVZ equation, at least, in the considered approximation.

\section*{Acknowledgements}
\hspace*{\parindent}

This work was supported by Foundation for Advancement of Theoretical Physics and Mathematics ``BASIS'', grant No. 19-1-1-45-1.

\appendix

\section*{Appendix}

\section{Contributions of various two-loop supergraphs to the $\beta$-function}
\hspace*{\parindent}\label{Appendix_Results_For_Supergraphs}

In this appendix we collect the expressions for contributions of the supergraphs B1 --- B7 to the function $\beta/\alpha_0^2$, which were obtained with the help of the method described in Sect. \ref{Section_Algorithm}. Certainly, by construction, the results are given by integrals of double total derivatives.

\begin{eqnarray}\label{Contribution_Gauge+Ghost}
&&\hspace*{-5mm} \Delta_{\mbox{\scriptsize gauge+ghost}}\Big(\frac{\beta}{\alpha_0^2}\Big) = 4\pi C_2^2 \frac{d}{d\ln\Lambda}\int \frac{d^4Q}{(2\pi)^4} \frac{d^4K}{(2\pi)^4}\, \left[\frac{\partial^2}{\partial Q_\mu^2} + \frac{\partial^2}{\partial K_\mu^2} - \frac{\partial^2}{\partial Q_\mu \partial K^\mu}\right] \frac{e_0^2}{R_K R_Q}\nonumber\\
&&\hspace*{-5mm} \times \left\{-\frac{R_K}{2 Q^2 K^2 (K+Q)^2}\right. - \frac{1}{2Q^2 K^2} \left(\frac{R_Q-R_K}{Q^2-K^2}\right)
- \frac{1}{R_{K+Q} K^2} \Big(1-\frac{Q^2}{2(K+Q)^2}\Big) \left(\frac{R_Q-R_K}{Q^2-K^2}\right)\nonumber\\
&&\hspace*{-4mm} \times\left(\frac{R_{K+Q}-R_Q}{(K+Q)^2-Q^2}\right) - \frac{1}{R_{K+Q} (K+Q)^2}\left(\frac{R_Q-R_K}{Q^2-K^2}\right)^2 +\frac{2}{K^2 \big((K+Q)^2-Q^2\big)^2}
\nonumber\\
&&\hspace*{-5mm} \times \left[ R_{K+Q} - R_Q - R_Q' \Big(\frac{(K+Q)^2}{\Lambda^2} - \frac{Q^2}{\Lambda^2}\Big)\right] - \frac{Q_\mu K^\mu}{Q^2 K^2} \left[\frac{R_{K+Q}}{\left((K+Q)^2-K^2\right)\left((K+Q)^2-Q^2\right)}\right. \nonumber\\
&&\hspace*{-5mm} \left. + \frac{R_{K}}{\left(K^2-(K+Q)^2\right)\left(K^2-Q^2\right)} \left. + \frac{R_{Q}}{\left(Q^2-(K+Q)^2\right)\left(Q^2-K^2\right)}\right]
\right\};\\
&&\vphantom{1}\nonumber\\
\label{Contribution_Varphi}
&&\hspace*{-5mm}  \Delta_\varphi\Big(\frac{\beta}{\alpha_0^2}\Big) = 4\pi C_2^2 \frac{d}{d\ln\Lambda}\int \frac{d^4Q}{(2\pi)^4} \frac{d^4K}{(2\pi)^4} \left[\frac{\partial^2}{\partial Q_\mu^2} + \frac{\partial^2}{\partial K_\mu^2} - \frac{\partial^2}{\partial Q_\mu\partial K^\mu}\right]\frac{e_0^2}{K^2 R_K}\left\{\vphantom{\frac{1}{2}}\right. \frac{1}{\big(Q^2+M_\varphi^2\big)}\nonumber\\
&&\hspace*{-5mm} \times
\frac{1}{\big((Q+K)^2+M_\varphi^2\big)} - \frac{1}{\big((K+Q)^2-Q^2\big)}\left[\vphantom{\frac{1}{2}}\right.
\frac{R_{K+Q}^2}{2\big((K+Q)^2 R_{K+Q}^2 + M_\varphi^2\big)} - \frac{R_{Q}^2}{2\big(Q^2 R_{Q}^2 + M_\varphi^2\big)} \nonumber\\
&&\hspace*{-5mm}  - \frac{M_\varphi^2 R_{K+Q}'}{\Lambda^2 R_{K+Q}\big((K+Q)^2 R_{K+Q}^2 + M_\varphi^2\big)} + \frac{M_\varphi^2 R_{Q}'}{\Lambda^2 R_{Q}\big(Q^2 R_{Q}^2 + M_\varphi^2\big)}\left.\vphantom{\frac{1}{2}} + \frac{R_{K+Q}'}{\Lambda^2 R_{K+Q}} - \frac{R_Q'}{\Lambda^2 R_Q}\right]
\left.\vphantom{\frac{1}{2}}\right\};\\
&&\vphantom{1}\nonumber\\
\label{Contribution_Matter}
&&\hspace*{-5mm}  \Delta_{\mbox{\scriptsize matter}}\Big(\frac{\beta}{\alpha_0^2}\Big) = \frac{4\pi}{r} \frac{d}{d\ln\Lambda}\int \frac{d^4Q}{(2\pi)^4} \frac{d^4K}{(2\pi)^4} \left[\mbox{tr}\left(C(R)^2\right)\frac{\partial^2}{\partial Q_\mu^2} + C_2 \mbox{tr}\, C(R)\Big(\frac{\partial^2}{\partial K_\mu^2} - \frac{\partial^2}{\partial Q_\mu\partial K^\mu}\Big)\right]\nonumber\\
&&\hspace*{-5mm} \times \frac{e_0^2}{K^2 R_K}\left\{\vphantom{\frac{1}{2}}\right.
\frac{1}{2Q^2 (Q+K)^2} + \frac{1}{\big((K+Q)^2-Q^2\big)}\left[\vphantom{\frac{1}{2}}\right.
\frac{F_{K+Q}^2}{2\big((K+Q)^2 F_{K+Q}^2 + M^2\big)} - \frac{F_{Q}^2}{2\big(Q^2 F_{Q}^2 + M^2\big)}\nonumber\\
&&\hspace*{-5mm} - \frac{M^2 F_{K+Q}'}{\Lambda^2 F_{K+Q}\big((K+Q)^2 F_{K+Q}^2 + M^2\big)} + \frac{M^2 F_{Q}'}{\Lambda^2 F_{Q}\big(Q^2 F_{Q}^2 + M^2\big)}\left.\vphantom{\frac{1}{2}} \right]
\left.\vphantom{\frac{1}{2}}\right\};\\
&&\vphantom{1}\nonumber\\
\label{Contribution_Yukawa}
&&\hspace*{-5mm}  \Delta_{\mbox{\scriptsize Yukawa}}\Big(\frac{\beta}{\alpha_0^2}\Big) = - \frac{2\pi}{r} \frac{d}{d\ln\Lambda} \int \frac{d^4Q}{(2\pi)^4} \frac{d^4K}{(2\pi)^4}\, \lambda_0^{ijk} \lambda^*_{0ijl} C(R)_k{}^l \frac{\partial^2}{\partial Q_\mu^2} \Big(\frac{1}{Q^2 F_Q K^2 F_K (Q+K)^2 F_{Q+K}}\Big).\nonumber\\
\end{eqnarray}


\begin{thebibliography}{100}

%\cite{Peskin:1995ev}
\bibitem{Peskin:1995ev}
  M.~E.~Peskin and D.~V.~Schroeder,
  ``An Introduction to quantum field theory,'' CRC Press, Taylor \& Francis group (2019) 866 p.
  %%CITATION = INSPIRE-407703;%%

%\cite{Mohapatra:1986uf}
\bibitem{Mohapatra:1986uf}
  R.~N.~Mohapatra,
  ``Unification And Supersymmetry. The Frontiers Of Quark - Lepton Physics : The Frontiers Of Quark-lepton Physics,''
  New York, USA: Springer (2003) 421~p.
  %doi:10.1007/978-1-4757-1928-4
  %%CITATION = doi:10.1007/978-1-4757-1928-4;%%

%\cite{tHooft:1972tcz}
\bibitem{tHooft:1972tcz}
  G.~'t Hooft and M.~J.~G.~Veltman,
  %``Regularization and Renormalization of Gauge Fields,''
  Nucl.\ Phys.\ B {\bf 44} (1972) 189.
  %doi:10.1016/0550-3213(72)90279-9
  %%CITATION = doi:10.1016/0550-3213(72)90279-9;%%

%\cite{Bollini:1972ui}
\bibitem{Bollini:1972ui}
  C.~G.~Bollini and J.~J.~Giambiagi,
  %``Dimensional Renormalization: The Number of Dimensions as a Regularizing Parameter,''
  Nuovo Cim.\ B {\bf 12} (1972) 20.
  %doi:10.1007/BF02895558
  %%CITATION = doi:10.1007/BF02895558;%%

%\cite{Ashmore:1972uj}
\bibitem{Ashmore:1972uj}
  J.~F.~Ashmore,
  %``A Method of Gauge Invariant Regularization,''
  Lett.\ Nuovo Cim.\  {\bf 4} (1972) 289.
  %doi:10.1007/BF02824407
  %%CITATION = doi:10.1007/BF02824407;%%

%\cite{Cicuta:1972jf}
\bibitem{Cicuta:1972jf}
  G.~M.~Cicuta and E.~Montaldi,
  %``Analytic renormalization via continuous space dimension,''
  Lett.\ Nuovo Cim.\  {\bf 4} (1972) 329.
  %doi:10.1007/BF02756527
  %%CITATION = doi:10.1007/BF02756527;%%

%\cite{Delbourgo:1974az}
\bibitem{Delbourgo:1974az}
  R.~Delbourgo and V.~B.~Prasad,
  %``Supersymmetry in the Four-Dimensional Limit,''
  J.\ Phys.\ G {\bf 1} (1975) 377.
  %doi:10.1088/0305-4616/1/4/001
  %%CITATION = doi:10.1088/0305-4616/1/4/001;%%

%\cite{Siegel:1979wq}
\bibitem{Siegel:1979wq}
  W.~Siegel,
  %``Supersymmetric Dimensional Regularization via Dimensional Reduction,''
  Phys.\ Lett.\  {\bf 84B} (1979) 193.
  %doi:10.1016/0370-2693(79)90282-X
  %%CITATION = doi:10.1016/0370-2693(79)90282-X;%%

%\cite{Siegel:1980qs}
\bibitem{Siegel:1980qs}
  W.~Siegel,
  %``Inconsistency of Supersymmetric Dimensional Regularization,''
  Phys.\ Lett.\  {\bf 94B} (1980) 37.
  %doi:10.1016/0370-2693(80)90819-9
  %%CITATION = doi:10.1016/0370-2693(80)90819-9;%%

%\cite{Avdeev:1981vf}
\bibitem{Avdeev:1981vf}
  L.~V.~Avdeev, G.~A.~Chochia and A.~A.~Vladimirov,
  %``On the Scope of Supersymmetric Dimensional Regularization,''
  Phys.\ Lett.\  {\bf 105B} (1981) 272.
  %doi:10.1016/0370-2693(81)90886-8
  %%CITATION = doi:10.1016/0370-2693(81)90886-8;%%

%\cite{Avdeev:1982xy}
\bibitem{Avdeev:1982xy}
  L.~V.~Avdeev and A.~A.~Vladimirov,
  %``Dimensional Regularization and Supersymmetry,''
  Nucl.\ Phys.\ B {\bf 219} (1983) 262.
  %doi:10.1016/0550-3213(83)90437-6
  %%CITATION = doi:10.1016/0550-3213(83)90437-6;%%

%\cite{Slavnov:1971aw}
\bibitem{Slavnov:1971aw}
  A.~A.~Slavnov,
  %``Invariant regularization of nonlinear chiral theories,''
  Nucl.\ Phys.\ B {\bf 31} (1971) 301.
  %doi:10.1016/0550-3213(71)90234-3
  %%CITATION = doi:10.1016/0550-3213(71)90234-3;%%

%\cite{Slavnov:1972sq}
\bibitem{Slavnov:1972sq}
  A.~A.~Slavnov,
  %``Invariant regularization of gauge theories,''
  Theor.Math.Phys. {\bf 13} (1972) 1064
   [Teor.\ Mat.\ Fiz.\  {\bf 13} (1972) 174].
  %%CITATION = TMFZA,13,174;%%

%\cite{Krivoshchekov:1978xg}
\bibitem{Krivoshchekov:1978xg}
  V.~K.~Krivoshchekov,
  %``Invariant Regularizations for Supersymmetric Gauge Theories,''
  Theor.\ Math.\ Phys.\ {\bf 36} (1978) 745
 [Teor.\ Mat.\ Fiz.\  {\bf 36} (1978) 291].
 %%CITATION = TMFZA,36,291;%%

%\cite{West:1985jx}
\bibitem{West:1985jx}
  P.~C.~West,
  %``Higher Derivative Regulation of Supersymmetric Theories,''
  Nucl.\ Phys.\ B {\bf 268} (1986) 113.
  %doi:10.1016/0550-3213(86)90203-8
  %%CITATION = doi:10.1016/0550-3213(86)90203-8;%%

%\cite{Buchbinder:2014wra}
\bibitem{Buchbinder:2014wra}
  I.~L.~Buchbinder and K.~V.~Stepanyantz,
  %``The higher derivative regularization and quantum corrections in N=2 supersymmetric theories,''
  Nucl.\ Phys.\ B {\bf 883} (2014) 20.
  %doi:10.1016/j.nuclphysb.2014.03.012
  %[arXiv:1402.5309 [hep-th]].
  %%CITATION = doi:10.1016/j.nuclphysb.2014.03.012;%%

%\cite{Galperin:1984av}
\bibitem{Galperin:1984av}
  A.~Galperin, E.~Ivanov, S.~Kalitzin, V.~Ogievetsky and E.~Sokatchev,
  %``Unconstrained N=2 matter, Yang-Mills and supergravity theories in harmonic superspace'',
  Class.\ Quant.\ Grav.\  {\bf 1} (1984) 469-498
   [Corrigendum ibid.\  {\bf 2} (1985) 127].
   %doi:10.1088/0264-9381/1/5/004.
  %%CITATION = CQGRD,1,469;%%

%\cite{Galperin:2001uw}
\bibitem{Galperin:2001uw}
  A.~S.~Galperin, E.~A.~Ivanov, V.~I.~Ogievetsky and E.~S.~Sokatchev,
  ``Harmonic superspace'',
  Cambridge, UK: Univ. Pr. (2001) 306 p.
  %doi:10.1017/CBO9780511535109.
  %%CITATION = doi:10.1017/CBO9780511535109;%%

%\cite{Buchbinder:2015eva}
\bibitem{Buchbinder:2015eva}
  I.~L.~Buchbinder, N.~G.~Pletnev and K.~V.~Stepanyantz,
  %``Manifestly N=2 supersymmetric regularization for N=2 supersymmetric field theories,''
  Phys.\ Lett.\ B {\bf 751} (2015) 434.
  %doi:10.1016/j.physletb.2015.10.071
  %[arXiv:1509.08055 [hep-th]].
  %%CITATION = doi:10.1016/j.physletb.2015.10.071;%%

%\cite{Grisaru:1982zh}
\bibitem{Grisaru:1982zh}
  M.~T.~Grisaru and W.~Siegel,
  %``Supergraphity. 2. Manifestly Covariant Rules and Higher Loop Finiteness,''
  Nucl.\ Phys.\ B {\bf 201} (1982) 292
   Erratum: [Nucl.\ Phys.\ B {\bf 206} (1982) 496].
  %doi:10.1016/0550-3213(82)90433-3, 10.1016/0550-3213(82)90282-6
  %%CITATION = doi:10.1016/0550-3213(82)90433-3, 10.1016/0550-3213(82)90282-6;%%

%\cite{Mandelstam:1982cb}
\bibitem{Mandelstam:1982cb}
  S.~Mandelstam,
  %``Light Cone Superspace and the Ultraviolet Finiteness of the N=4 Model,''
  Nucl.\ Phys.\ B {\bf 213} (1983) 149.
  %doi:10.1016/0550-3213(83)90179-7
  %%CITATION = doi:10.1016/0550-3213(83)90179-7;%%

%\cite{Brink:1982pd}
\bibitem{Brink:1982pd}
  L.~Brink, O.~Lindgren and B.~E.~W.~Nilsson,
  %``N=4 Yang-Mills Theory on the Light Cone,''
  Nucl.\ Phys.\ B {\bf 212} (1983) 401.
  %doi:10.1016/0550-3213(83)90678-8
  %%CITATION = doi:10.1016/0550-3213(83)90678-8;%%

%\cite{Howe:1983sr}
\bibitem{Howe:1983sr}
  P.~S.~Howe, K.~S.~Stelle and P.~K.~Townsend,
  %``Miraculous Ultraviolet Cancellations in Supersymmetry Made Manifest,''
  Nucl.\ Phys.\ B {\bf 236} (1984) 125.
  %doi:10.1016/0550-3213(84)90528-5
  %%CITATION = doi:10.1016/0550-3213(84)90528-5;%%

%\cite{Buchbinder:1997ib}
\bibitem{Buchbinder:1997ib}
  I.~L.~Buchbinder, S.~M.~Kuzenko and B.~A.~Ovrut,
  %``On the D = 4, N=2 nonrenormalization theorem,''
  Phys.\ Lett.\ B {\bf 433} (1998) 335.
  %doi:10.1016/S0370-2693(98)00688-1
  %[hep-th/9710142].
  %%CITATION = doi:10.1016/S0370-2693(98)00688-1;%%

%\cite{Faddeev:1980be}
\bibitem{Faddeev:1980be}
  L.~D.~Faddeev and A.~A.~Slavnov,
  %``Gauge Fields. Introduction To Quantum Theory,''
  Front.\ Phys.\  {\bf 50} (1980) 1
   [Front.\ Phys.\  (1991) 1].
  %%CITATION = FRPHA,50,1;%%

%\cite{Slavnov:1977zf}
\bibitem{Slavnov:1977zf}
  A.~A.~Slavnov,
  %``The Pauli-Villars Regularization for Nonabelian Gauge Theories,''
  Theor.\ Math.\ Phys.\  {\bf 33} (1977) 977
   [Teor.\ Mat.\ Fiz.\  {\bf 33} (1977) 210].
  %doi:10.1007/BF01036595
  %%CITATION = doi:10.1007/BF01036595;%%

%\cite{DeWitt:1965jb}
\bibitem{DeWitt:1965jb}
  B.~S.~DeWitt,
  ``Dynamical theory of groups and fields,''
  %Conf.\ Proc.\ C {\bf 630701} (1964) 585
  % [Les Houches Lect.\ Notes {\bf 13} (1964) 585]
  Gordon and Breach, New York, 1965.

%\cite{Abbott:1980hw}
\bibitem{Abbott:1980hw}
  L.~F.~Abbott,
  %``The Background Field Method Beyond One Loop,''
  Nucl.\ Phys.\ B {\bf 185} (1981) 189.
  %doi:10.1016/0550-3213(81)90371-0
  %%CITATION = doi:10.1016/0550-3213(81)90371-0;%%

%\cite{Abbott:1981ke}
\bibitem{Abbott:1981ke}
  L.~F.~Abbott,
  %``Introduction to the Background Field Method,''
  Acta Phys.\ Polon.\ B {\bf 13} (1982) 33.
  %%CITATION = APPOA,B13,33;%%

%\cite{Gates:1983nr}
\bibitem{Gates:1983nr}
  S.~J.~Gates, M.~T.~Grisaru, M.~Rocek and W.~Siegel,
  %``Superspace Or One Thousand and One Lessons in Supersymmetry,''
  Front.\ Phys.\  {\bf 58} (1983) 1.
  %[hep-th/0108200].
  %%CITATION = HEP-TH/0108200;%%

%\cite{Piguet:1981fb}
\bibitem{Piguet:1981fb}
  O.~Piguet and K.~Sibold,
  %``Renormalization of $N=1$ Supersymmetrical {Yang-Mills} Theories. 1. The Classical Theory,''
  Nucl.\ Phys.\ B {\bf 197} (1982) 257.
  %doi:10.1016/0550-3213(82)90291-7
  %%CITATION = doi:10.1016/0550-3213(82)90291-7;%%

%\cite{Piguet:1981hh}
\bibitem{Piguet:1981hh}
  O.~Piguet and K.~Sibold,
  %``Renormalization of $N=1$ Supersymmetrical {Yang-Mills} Theories. 2. The Radiative Corrections,''
  Nucl.\ Phys.\ B {\bf 197} (1982) 272.
  %doi:10.1016/0550-3213(82)90292-9
  %%CITATION = doi:10.1016/0550-3213(82)90292-9;%%

%\cite{Tyutin:1983rg}
\bibitem{Tyutin:1983rg}
  I.~V.~Tyutin,
  %``Renormalization Of Supergauge Theories With Nonextended Supersymmetry. (in Russian),''
  Yad.\ Fiz.\  {\bf 37} (1983) 761.
  %%CITATION = YAFIA,37,761;%%

%\cite{Juer:1982fb}
\bibitem{Juer:1982fb}
  J.~W.~Juer and D.~Storey,
  %``Nonlinear Renormalization in Superfield Gauge Theories,''
  Phys.\ Lett.\  {\bf 119B} (1982) 125.
  %doi:10.1016/0370-2693(82)90259-3
  %%CITATION = doi:10.1016/0370-2693(82)90259-3;%%

%\cite{Juer:1982mp}
\bibitem{Juer:1982mp}
  J.~W.~Juer and D.~Storey,
  %``One Loop Renormalization of Superfield {Yang-Mills} Theories,''
  Nucl.\ Phys.\ B {\bf 216} (1983) 185.
  %doi:10.1016/0550-3213(83)90491-1
  %%CITATION = doi:10.1016/0550-3213(83)90491-1;%%

%\cite{Kazantsev:2018kjx}
\bibitem{Kazantsev:2018kjx}
  A.~E.~Kazantsev, M.~D.~Kuzmichev, N.~P.~Meshcheriakov, S.~V.~Novgorodtsev, I.~E.~Shirokov, M.~B.~Skoptsov and K.~V.~Stepanyantz,
  %``Two-loop renormalization of the Faddeev-Popov ghosts in $ \mathcal{N}=1 $ supersymmetric gauge theories regularized by higher derivatives,''
  JHEP {\bf 1806} (2018) 020.
  %doi:10.1007/JHEP06(2018)020
  %[arXiv:1805.03686 [hep-th]].
  %%CITATION = doi:10.1007/JHEP06(2018)020;%%

%\cite{Becchi:1974md}
\bibitem{Becchi:1974md}
  C.~Becchi, A.~Rouet and R.~Stora,
  %``Renormalization of the Abelian Higgs-Kibble Model,''
  Commun.\ Math.\ Phys.\  {\bf 42} (1975) 127.
  %doi:10.1007/BF01614158
  %%CITATION = doi:10.1007/BF01614158;%%

%\cite{Tyutin:1975qk}
\bibitem{Tyutin:1975qk}
  I.~V.~Tyutin,
  ``Gauge Invariance in Field Theory and Statistical Physics in Operator Formalism,''
  arXiv:0812.0580 [hep-th].
  %%CITATION = ARXIV:0812.0580;%%

%\cite{Kazantsev:2017fdc}
\bibitem{Kazantsev:2017fdc}
  A.~E.~Kazantsev, M.~B.~Skoptsov and K.~V.~Stepanyantz,
  %``One-loop polarization operator of the quantum gauge superfield for ${\cal N}=1$ SYM regularized by higher derivatives,''
  Mod.\ Phys.\ Lett.\ A {\bf 32} (2017) no.36,  1750194.
  %doi:10.1142/S0217732317501942
  %[arXiv:1709.08575 [hep-th]].
  %%CITATION = doi:10.1142/S0217732317501942;%%

%\cite{Soloshenko:2003nc}
\bibitem{Soloshenko:2003nc}
  A.~A.~Soloshenko and K.~V.~Stepanyantz,
  %``Three loop beta function for N=1 supersymmetric electrodynamics, regularized by higher derivatives,''
  Theor.\ Math.\ Phys.\  {\bf 140} (2004) 1264
   [Teor.\ Mat.\ Fiz.\  {\bf 140} (2004) 437].
  %doi:10.1023/B:TAMP.0000039832.82367.50
  %[hep-th/0304083].
  %%CITATION = doi:10.1023/B:TAMP.0000039832.82367.50;%%

%\cite{Smilga:2004zr}
\bibitem{Smilga:2004zr}
  A.~V.~Smilga and A.~Vainshtein,
  %``Background field calculations and nonrenormalization theorems in 4-D supersymmetric gauge theories and their low-dimensional descendants,''
  Nucl.\ Phys.\ B {\bf 704} (2005) 445.
  %doi:10.1016/j.nuclphysb.2004.10.010
  %[hep-th/0405142].
  %%CITATION = doi:10.1016/j.nuclphysb.2004.10.010;%%

%\cite{Stepanyantz:2011jy}
\bibitem{Stepanyantz:2011jy}
  K.~V.~Stepanyantz,
  %``Derivation of the exact NSVZ $\beta$-function in N=1 SQED, regularized by higher derivatives, by direct summation of Feynman diagrams,''
  Nucl.\ Phys.\ B {\bf 852} (2011) 71.
  %doi:10.1016/j.nuclphysb.2011.06.018
  %[arXiv:1102.3772 [hep-th]].
  %%CITATION = doi:10.1016/j.nuclphysb.2011.06.018;%%

%\cite{Stepanyantz:2014ima}
\bibitem{Stepanyantz:2014ima}
  K.~V.~Stepanyantz,
  %``The NSVZ $\beta$-function and the Schwinger-Dyson equations for $\mathcal{N}=1$ SQED with $N_{f}$ flavors, regularized by higher derivatives,''
  JHEP {\bf 1408} (2014) 096.
  %doi:10.1007/JHEP08(2014)096
  %[arXiv:1404.6717 [hep-th]].
  %%CITATION = doi:10.1007/JHEP08(2014)096;%%

%\cite{Kazantsev:2014yna}
\bibitem{Kazantsev:2014yna}
  A.~E.~Kazantsev and K.~V.~Stepanyantz,
  %``Relation between two-point Green’s functions of $\mathcal{N} = 1$ SQED with N$_{f}$ flavors, regularized by higher derivatives, in the three-loop approximation,''
  J.\ Exp.\ Theor.\ Phys.\  {\bf 120} (2015) no.4,  618
   [Zh.\ Eksp.\ Teor.\ Fiz.\  {\bf 147} (2015) no.4,  714].
  %doi:10.1134/S1063776115040068
  %[arXiv:1410.1133 [hep-th]].
  %%CITATION = doi:10.1134/S1063776115040068;%%

%\cite{Vainshtein:1986ja}
\bibitem{Vainshtein:1986ja}
  A.~I.~Vainshtein, V.~I.~Zakharov and M.~A.~Shifman,
  %``Gell-mann-low Function In Supersymmetric Electrodynamics,''
  JETP Lett.\  {\bf 42} (1985) 224
   [Pisma Zh.\ Eksp.\ Teor.\ Fiz.\  {\bf 42} (1985) 182].
  %%CITATION = JTPLA,42,224;%%

%\cite{Shifman:1985fi}
\bibitem{Shifman:1985fi}
  M.~A.~Shifman, A.~I.~Vainshtein and V.~I.~Zakharov,
  %``Exact Gell-mann-low Function In Supersymmetric Electrodynamics,''
  Phys.\ Lett.\  {\bf 166B} (1986) 334.
  %doi:10.1016/0370-2693(86)90811-7
  %%CITATION = doi:10.1016/0370-2693(86)90811-7;%%

%\cite{Aleshin:2015qqc}
\bibitem{Aleshin:2015qqc}
  S.~S.~Aleshin, A.~L.~Kataev and K.~V.~Stepanyantz,
  %``Structure of three-loop contributions to the $\beta$-function of $\mathcal N = 1$ supersymmetric QED with N$_{f}$ flavors regularized by the dimensional reduction,''
  JETP Lett.\  {\bf 103} (2016) no.2,  77.
  %doi:10.1134/S0021364016020028
  %[arXiv:1511.05675 [hep-th]].
  %%CITATION = doi:10.1134/S0021364016020028;%%

%\cite{Aleshin:2016rrr}
\bibitem{Aleshin:2016rrr}
  S.~S.~Aleshin, I.~O.~Goriachuk, A.~L.~Kataev and K.~V.~Stepanyantz,
  %``The NSVZ scheme for ${\cal N}=1$ SQED with $N_f$ flavors, regularized by the dimensional reduction, in the three-loop approximation,''
  Phys.\ Lett.\ B {\bf 764} (2017) 222.
  %doi:10.1016/j.physletb.2016.11.041
  %[arXiv:1610.08034 [hep-th]].
  %%CITATION = doi:10.1016/j.physletb.2016.11.041;%%

%\cite{Goriachuk:2018cac}
\bibitem{Goriachuk:2018cac}
  I.~O.~Goriachuk, A.~L.~Kataev and K.~V.~Stepanyantz,
  %``A class of the NSVZ renormalization schemes for ${\cal N}=1$ SQED,''
  Phys.\ Lett.\ B {\bf 785} (2018) 561.
  %doi:10.1016/j.physletb.2018.09.014
  %[arXiv:1808.02050 [hep-th]].
  %%CITATION = doi:10.1016/j.physletb.2018.09.014;%%

%\cite{Bardeen:1978yd}
\bibitem{Bardeen:1978yd}
  W.~A.~Bardeen, A.~J.~Buras, D.~W.~Duke and T.~Muta,
  %``Deep Inelastic Scattering Beyond the Leading Order in Asymptotically Free Gauge Theories,''
  Phys.\ Rev.\ D {\bf 18} (1978) 3998.
  %doi:10.1103/PhysRevD.18.3998
  %%CITATION = doi:10.1103/PhysRevD.18.3998;%%

%\cite{Jack:1996vg}
\bibitem{Jack:1996vg}
  I.~Jack, D.~R.~T.~Jones and C.~G.~North,
  %``N=1 supersymmetry and the three loop gauge Beta function,''
  Phys.\ Lett.\ B {\bf 386} (1996) 138.
  %doi:10.1016/0370-2693(96)00918-5
  %[hep-ph/9606323].
  %%CITATION = doi:10.1016/0370-2693(96)00918-5;%%

%\cite{Jack:1996cn}
\bibitem{Jack:1996cn}
  I.~Jack, D.~R.~T.~Jones and C.~G.~North,
  %``Scheme dependence and the NSVZ Beta function,''
  Nucl.\ Phys.\ B {\bf 486} (1997) 479.
  %doi:10.1016/S0550-3213(96)00637-2
  %[hep-ph/9609325].
  %%CITATION = doi:10.1016/S0550-3213(96)00637-2;%%

%\cite{Jack:1998uj}
\bibitem{Jack:1998uj}
  I.~Jack, D.~R.~T.~Jones and A.~Pickering,
  %``The Connection between DRED and NSVZ,''
  Phys.\ Lett.\ B {\bf 435} (1998) 61.
  %doi:10.1016/S0370-2693(98)00769-2
  %[hep-ph/9805482].
  %%CITATION = doi:10.1016/S0370-2693(98)00769-2;%%

%\cite{Harlander:2006xq}
\bibitem{Harlander:2006xq}
  R.~V.~Harlander, D.~R.~T.~Jones, P.~Kant, L.~Mihaila and M.~Steinhauser,
  %``Four-loop beta function and mass anomalous dimension in dimensional reduction,''
  JHEP {\bf 0612} (2006) 024.
  %doi:10.1088/1126-6708/2006/12/024
  %[hep-ph/0610206].
  %%CITATION = doi:10.1088/1126-6708/2006/12/024;%%

%\cite{Mihaila:2013wma}
\bibitem{Mihaila:2013wma}
  L.~Mihaila,
  %``Precision Calculations in Supersymmetric Theories,''
  Adv.\ High Energy Phys.\  {\bf 2013} (2013) 607807.
  %doi:10.1155/2013/607807
  %[arXiv:1310.6178 [hep-ph]].
  %%CITATION = doi:10.1155/2013/607807;%%

%\cite{Kataev:2013eta}
\bibitem{Kataev:2013eta}
  A.~L.~Kataev and K.~V.~Stepanyantz,
  %``NSVZ scheme with the higher derivative regularization for $\mathcal{N} =$ 1 SQED,''
  Nucl.\ Phys.\ B {\bf 875} (2013) 459.
  %doi:10.1016/j.nuclphysb.2013.07.010
  %[arXiv:1305.7094 [hep-th]].
  %%CITATION = doi:10.1016/j.nuclphysb.2013.07.010;%%

%\cite{Kataev:2013csa}
\bibitem{Kataev:2013csa}
  A.~L.~Kataev and K.~V.~Stepanyantz,
  %``Scheme independent consequence of the NSVZ relation for N=1 SQED with N_f flavors,''
  Phys.\ Lett.\ B {\bf 730} (2014) 184.
  %doi:10.1016/j.physletb.2014.01.053
  %[arXiv:1311.0589 [hep-th]].
  %%CITATION = doi:10.1016/j.physletb.2014.01.053;%%

%\cite{Kataev:2014gxa}
\bibitem{Kataev:2014gxa}
  A.~L.~Kataev and K.~V.~Stepanyantz,
  %``The NSVZ beta-function in supersymmetric theories with different regularizations and renormalization prescriptions,''
  Theor.\ Math.\ Phys.\  {\bf 181} (2014) 1531.
  %doi:10.1007/s11232-014-0233-3
  %[arXiv:1405.7598 [hep-th]].
  %%CITATION = doi:10.1007/s11232-014-0233-3;%%

%\cite{Shakhmanov:2017wji}
\bibitem{Shakhmanov:2017wji}
  V.~Y.~Shakhmanov and K.~V.~Stepanyantz,
  %``New form of the NSVZ relation at the two-loop level,''
  Phys.\ Lett.\ B {\bf 776} (2018) 417.
  %doi:10.1016/j.physletb.2017.12.005
  %[arXiv:1711.03899 [hep-th]].
  %%CITATION = doi:10.1016/j.physletb.2017.12.005;%%

%\cite{Stepanyantz:2017sqg}
\bibitem{Stepanyantz:2017sqg}
  K.~V.~Stepanyantz,
  %``Structure of Quantum Corrections in ${\cal N}=1$ Supersymmetric Gauge Theories,''
  Bled Workshops Phys.\  {\bf 18} (2017) no.2,  197.
  %[arXiv:1711.09194 [hep-th]].
  %%CITATION = ARXIV:1711.09194;%%

%\cite{Kataev:2019olb}
\bibitem{Kataev:2019olb}
  A.~L.~Kataev, A.~E.~Kazantsev and K.~V.~Stepanyantz,
  %``On-shell renormalization scheme for ${\cal N}=1$ SQED and the NSVZ relation,''
  Eur.\ Phys.\ J.\ C {\bf 79} (2019) no.6,  477.
  %doi:10.1140/epjc/s10052-019-6993-z
  %[arXiv:1905.02222 [hep-th]].
  %%CITATION = doi:10.1140/epjc/s10052-019-6993-z;%%

%\cite{Hisano:1997ua}
\bibitem{Hisano:1997ua}
  J.~Hisano and M.~A.~Shifman,
  %``Exact results for soft supersymmetry breaking parameters in supersymmetric gauge theories,''
  Phys.\ Rev.\ D {\bf 56} (1997) 5475.
  %doi:10.1103/PhysRevD.56.5475
  %[hep-ph/9705417].
  %%CITATION = doi:10.1103/PhysRevD.56.5475;%%

%\cite{Jack:1997pa}
\bibitem{Jack:1997pa}
  I.~Jack and D.~R.~T.~Jones,
  %``The Gaugino Beta function,''
  Phys.\ Lett.\ B {\bf 415} (1997) 383.
  %doi:10.1016/S0370-2693(97)01277-X
  %[hep-ph/9709364].
  %%CITATION = doi:10.1016/S0370-2693(97)01277-X;%%

%\cite{Avdeev:1997vx}
\bibitem{Avdeev:1997vx}
  L.~V.~Avdeev, D.~I.~Kazakov and I.~N.~Kondrashuk,
  %``Renormalizations in softly broken SUSY gauge theories,''
  Nucl.\ Phys.\ B {\bf 510} (1998) 289.
  %doi:10.1016/S0550-3213(98)81015-8, 10.1016/S0550-3213(97)00706-2
  %[hep-ph/9709397].
  %%CITATION = doi:10.1016/S0550-3213(98)81015-8, 10.1016/S0550-3213(97)00706-2;%%

%\cite{Nartsev:2016nym}
\bibitem{Nartsev:2016nym}
  I.~V.~Nartsev and K.~V.~Stepanyantz,
  %``Exact renormalization of the photino mass in softly broken $ \mathcal{N} $ = 1 SQED with N$_{f}$ flavors regularized by higher derivatives,''
  JHEP {\bf 1704} (2017) 047.
  %doi:10.1007/JHEP04(2017)047
  %[arXiv:1610.01280 [hep-th]].
  %%CITATION = doi:10.1007/JHEP04(2017)047;%%

%\cite{Shifman:2014cya}
\bibitem{Shifman:2014cya}
  M.~Shifman and K.~Stepanyantz,
  %``Exact Adler Function in Supersymmetric QCD,''
  Phys.\ Rev.\ Lett.\  {\bf 114} (2015) no.5,  051601.
  %doi:10.1103/PhysRevLett.114.051601
  %[arXiv:1412.3382 [hep-th]].
  %%CITATION = doi:10.1103/PhysRevLett.114.051601;%%

%\cite{Shifman:2015doa}
\bibitem{Shifman:2015doa}
  M.~Shifman and K.~V.~Stepanyantz,
  %``Derivation of the exact expression for the D function in N=1 SQCD,''
  Phys.\ Rev.\ D {\bf 91} (2015) 105008.
  %doi:10.1103/PhysRevD.91.105008
  %[arXiv:1502.06655 [hep-th]].
  %%CITATION = doi:10.1103/PhysRevD.91.105008;%%

%\cite{Nartsev:2016mvn}
\bibitem{Nartsev:2016mvn}
  I.~V.~Nartsev and K.~V.~Stepanyantz,
  %``NSVZ-like scheme for the photino mass in softly broken ${\cal N}=1$ SQED regularized by higher derivatives,''
  JETP Lett.\  {\bf 105} (2017) no.2,  69.
  %doi:10.1134/S0021364017020059
  %[arXiv:1611.09091 [hep-th]].
  %%CITATION = doi:10.1134/S0021364017020059;%%

%\cite{Kataev:2017qvk}
\bibitem{Kataev:2017qvk}
  A.~L.~Kataev, A.~E.~Kazantsev and K.~V.~Stepanyantz,
  %``The Adler $D$-function for ${\cal N}=1$ SQCD regularized by higher covariant derivatives in the three-loop approximation,''
  Nucl.\ Phys.\ B {\bf 926} (2018) 295.
  %doi:10.1016/j.nuclphysb.2017.11.009
  %[arXiv:1710.03941 [hep-th]].
  %%CITATION = doi:10.1016/j.nuclphysb.2017.11.009;%%

%\cite{Aleshin:2019yqj}
\bibitem{Aleshin:2019yqj}
  S.~S.~Aleshin, A.~L.~Kataev and K.~V.~Stepanyantz,
  %``The three-loop Adler $D$-function for ${\cal N}=1$ SQCD regularized by dimensional reduction,''
  JHEP {\bf 1903} (2019) 196.
  %doi:10.1007/JHEP03(2019)196
  %[arXiv:1902.08602 [hep-th]].
  %%CITATION = doi:10.1007/JHEP03(2019)196;%%

%\cite{Pimenov:2009hv}
\bibitem{Pimenov:2009hv}
  A.~B.~Pimenov, E.~S.~Shevtsova and K.~V.~Stepanyantz,
  %``Calculation of two-loop beta-function for general N=1 supersymmetric Yang--Mills theory with the higher covariant derivative regularization,''
  Phys.\ Lett.\ B {\bf 686} (2010) 293.
  %doi:10.1016/j.physletb.2010.02.047
  %[arXiv:0912.5191 [hep-th]].
  %%CITATION = doi:10.1016/j.physletb.2010.02.047;%%

%\cite{Stepanyantz:2011cpt}
\bibitem{Stepanyantz:2011cpt}
  K.~V.~Stepanyantz,
  %``Higher covariant derivative regularization for calculations in supersymmetric theories,''
  Proc.\ Steklov Inst.\ Math.\  {\bf 272} (2011) no.1,  256.
  %doi:10.1134/S008154381101024X
  %%CITATION = doi:10.1134/S008154381101024X;%%

%\cite{Stepanyantz:2011zz}
\bibitem{Stepanyantz:2011zz}
  K.~V.~Stepanyantz,
  %``Quantum corrections in N=1 supersymmetric theories with cubic superpotential, regularized by higher covariant derivatives,''
  Phys.\ Part.\ Nucl.\ Lett.\  {\bf 8} (2011) 321.
  %doi:10.1134/S1547477111030198
  %%CITATION = doi:10.1134/S1547477111030198;%%

%\cite{Stepanyantz:2011bz}
\bibitem{Stepanyantz:2011bz}
  K.~V.~Stepanyantz,
  ``Factorization of integrals defining the two-loop $\beta$-function for the general renormalizable N=1 SYM theory, regularized by the higher covariant derivatives, into integrals of double total derivatives,''
  arXiv:1108.1491 [hep-th].
  %%CITATION = ARXIV:1108.1491;%%

%\cite{Aleshin:2016yvj}
\bibitem{Aleshin:2016yvj}
  S.~S.~Aleshin, A.~E.~Kazantsev, M.~B.~Skoptsov and K.~V.~Stepanyantz,
  %``One-loop divergences in non-Abelian supersymmetric theories regularized by BRST-invariant version of the higher derivative regularization,''
  JHEP {\bf 1605} (2016) 014.
  %doi:10.1007/JHEP05(2016)014
  %[arXiv:1603.04347 [hep-th]].
  %%CITATION = doi:10.1007/JHEP05(2016)014;%%

%\cite{Shakhmanov:2017soc}
\bibitem{Shakhmanov:2017soc}
  V.~Y.~Shakhmanov and K.~V.~Stepanyantz,
  %``Three-loop NSVZ relation for terms quartic in the Yukawa couplings with the higher covariant derivative regularization,''
  Nucl.\ Phys.\ B {\bf 920} (2017) 345.
  %doi:10.1016/j.nuclphysb.2017.04.017
  %[arXiv:1703.10569 [hep-th]].
  %%CITATION = doi:10.1016/j.nuclphysb.2017.04.017;%%

%\cite{Kazantsev:2018nbl}
\bibitem{Kazantsev:2018nbl}
  A.~E.~Kazantsev, V.~Y.~Shakhmanov and K.~V.~Stepanyantz,
  %``New form of the exact NSVZ $\beta$-function: the three-loop verification for terms containing Yukawa couplings,''
  JHEP {\bf 1804} (2018) 130.
  %doi:10.1007/JHEP04(2018)130
  %[arXiv:1803.06612 [hep-th]].
  %%CITATION = doi:10.1007/JHEP04(2018)130;%%

%\cite{Stepanyantz:2019ihw}
\bibitem{Stepanyantz:2019ihw}
  K.~V.~Stepanyantz,
  %``The $\beta$-function of ${\cal N}=1$ supersymmetric gauge theories regularized by higher covariant derivatives as an integral of double total derivatives,''
  JHEP {\bf 1910} (2019) 011.
  %doi:10.1007/JHEP10(2019)011
  %[arXiv:1908.04108 [hep-th]].
  %%CITATION = doi:10.1007/JHEP10(2019)011;%%

%\cite{Novikov:1983uc}
\bibitem{Novikov:1983uc}
  V.~A.~Novikov, M.~A.~Shifman, A.~I.~Vainshtein and V.~I.~Zakharov,
  %``Exact Gell-Mann-Low Function of Supersymmetric Yang-Mills Theories from Instanton Calculus,''
  Nucl.\ Phys.\ B {\bf 229} (1983) 381.
  %doi:10.1016/0550-3213(83)90338-3
  %%CITATION = doi:10.1016/0550-3213(83)90338-3;%%

%\cite{Jones:1983ip}
\bibitem{Jones:1983ip}
  D.~R.~T.~Jones,
  %``More on the Axial Anomaly in Supersymmetric {Yang-Mills} Theory,''
  Phys.\ Lett.\  {\bf 123B} (1983) 45.
  %doi:10.1016/0370-2693(83)90955-3
  %%CITATION = doi:10.1016/0370-2693(83)90955-3;%%

%\cite{Novikov:1985rd}
\bibitem{Novikov:1985rd}
  V.~A.~Novikov, M.~A.~Shifman, A.~I.~Vainshtein and V.~I.~Zakharov,
  %``Beta Function in Supersymmetric Gauge Theories: Instantons Versus Traditional Approach,''
  Phys.\ Lett.\  {\bf 166B} (1986) 329
   [Sov.\ J.\ Nucl.\ Phys.\  {\bf 43} (1986) 294]
   [Yad.\ Fiz.\  {\bf 43} (1986) 459].
  %doi:10.1016/0370-2693(86)90810-5
  %%CITATION = doi:10.1016/0370-2693(86)90810-5;%%

%\cite{Shifman:1986zi}
\bibitem{Shifman:1986zi}
  M.~A.~Shifman and A.~I.~Vainshtein,
  %``Solution of the Anomaly Puzzle in SUSY Gauge Theories and the Wilson Operator Expansion,''
  Nucl.\ Phys.\ B {\bf 277} (1986) 456
   [Sov.\ Phys.\ JETP {\bf 64} (1986) 428]
   [Zh.\ Eksp.\ Teor.\ Fiz.\  {\bf 91} (1986) 723].
  %doi:10.1016/0550-3213(86)90451-7
  %%CITATION = doi:10.1016/0550-3213(86)90451-7;%%

%\cite{Stepanyantz:2016gtk}
\bibitem{Stepanyantz:2016gtk}
  K.~V.~Stepanyantz,
  %``Non-renormalization of the $V\bar cc$-vertices in ${\cal N}=1$ supersymmetric theories,''
  Nucl.\ Phys.\ B {\bf 909} (2016) 316.
  %doi:10.1016/j.nuclphysb.2016.05.011
  %[arXiv:1603.04801 [hep-th]].
  %%CITATION = doi:10.1016/j.nuclphysb.2016.05.011;%%

%\cite{Taylor:1971ff}
\bibitem{Taylor:1971ff}
  J.~C.~Taylor,
  %``Ward Identities and Charge Renormalization of the Yang-Mills Field,''
  Nucl.\ Phys.\ B {\bf 33} (1971) 436.
  %doi:10.1016/0550-3213(71)90297-5
  %%CITATION = doi:10.1016/0550-3213(71)90297-5;%%

%\cite{Slavnov:1972fg}
\bibitem{Slavnov:1972fg}
  A.~A.~Slavnov,
  %``Ward Identities in Gauge Theories,''
  Theor.\ Math.\ Phys.\  {\bf 10} (1972) 99
   [Teor.\ Mat.\ Fiz.\  {\bf 10} (1972) 153].
  %doi:10.1007/BF01090719
  %%CITATION = doi:10.1007/BF01090719;%%

%\cite{West:1990tg}
\bibitem{West:1990tg}
  P.~C.~West,
  ``Introduction to supersymmetry and supergravity,''
  Singapore, Singapore: World Scientific (1990) 425 p.

%\cite{Buchbinder:1998qv}
\bibitem{Buchbinder:1998qv}
  I.~L.~Buchbinder and S.~M.~Kuzenko,
  ``Ideas and methods of supersymmetry and supergravity: Or a walk through superspace,''
  Bristol, UK: IOP (1998) 656 p.

%\cite{Dudal:2002pq}
\bibitem{Dudal:2002pq}
  D.~Dudal, H.~Verschelde and S.~P.~Sorella,
  %``The Anomalous dimension of the composite operator A**2 in the Landau gauge,''
  Phys.\ Lett.\ B {\bf 555} (2003) 126.
  %doi:10.1016/S0370-2693(03)00043-1
  %[hep-th/0212182].
  %%CITATION = doi:10.1016/S0370-2693(03)00043-1;%%

%\cite{Capri:2014jqa}
\bibitem{Capri:2014jqa}
  M.~A.~L.~Capri, D.~R.~Granado, M.~S.~Guimaraes, I.~F.~Justo, L.~Mihaila, S.~P.~Sorella and D.~Vercauteren,
  %``Renormalization aspects of N=1 Super Yang-Mills theory in the Wess-Zumino gauge,''
  Eur.\ Phys.\ J.\ C {\bf 74} (2014) no.4,  2844.
  %doi:10.1140/epjc/s10052-014-2844-0
  %[arXiv:1401.6303 [hep-th]].
  %%CITATION = doi:10.1140/epjc/s10052-014-2844-0;%%

%\cite{Kuzmichev:2019ywn}
\bibitem{Kuzmichev:2019ywn}
  M.~D.~Kuzmichev, N.~P.~Meshcheriakov, S.~V.~Novgorodtsev, I.~E.~Shirokov and K.~V.~Stepanyantz,
  %``Three-loop contribution of the Faddeev-Popov ghosts to the $\beta$-function of ${\cal N}=1$ supersymmetric gauge theories and the NSVZ relation,''
  Eur.\ Phys.\ J.\ C {\bf 79} (2019) no.9,  809.
  %doi:10.1140/epjc/s10052-019-7323-1
  %[arXiv:1908.10586 [hep-th]].
  %%CITATION = doi:10.1140/epjc/s10052-019-7323-1;%%

%\cite{Slavnov:2001pu}
\bibitem{Slavnov:2001pu}
  A.~A.~Slavnov,
  %``Universal gauge invariant renormalization,''
  Phys.\ Lett.\ B {\bf 518} (2001) 195.
  %doi:10.1016/S0370-2693(01)01002-4
  %%CITATION = doi:10.1016/S0370-2693(01)01002-4;%%

%\cite{Slavnov:2002ir}
\bibitem{Slavnov:2002ir}
  A.~A.~Slavnov,
  %``Regularization-independent gauge-invariant renormalization of the Yang-Mills theory,''
  Theor.\ Math.\ Phys.\  {\bf 130} (2002) 1
   [Teor.\ Mat.\ Fiz.\  {\bf 130} (2002) 3].
  %doi:10.1023/A:1013828529525
  %%CITATION = doi:10.1023/A:1013828529525;%%

%\cite{Slavnov:2002kg}
\bibitem{Slavnov:2002kg}
  A.~A.~Slavnov and K.~V.~Stepanyantz,
  %``Universal invariant renormalization for supersymmetric theories,''
  Theor.\ Math.\ Phys.\  {\bf 135} (2003) 673
   [Teor.\ Mat.\ Fiz.\  {\bf 135} (2003) 265].
  %doi:10.1023/A:1023622616220
  %[hep-th/0208006].
  %%CITATION = doi:10.1023/A:1023622616220;%%

%\cite{Slavnov:2003cx}
\bibitem{Slavnov:2003cx}
  A.~A.~Slavnov and K.~V.~Stepanyantz,
  %``Universal invariant renormalization of supersymmetric Yang-Mills theory,''
  Theor.\ Math.\ Phys.\  {\bf 139} (2004) 599
   [Teor.\ Mat.\ Fiz.\  {\bf 139} (2004) 179].
  %doi:10.1023/B:TAMP.0000026178.67671.6a
  %[hep-th/0305128].
  %%CITATION = doi:10.1023/B:TAMP.0000026178.67671.6a;%%

%\cite{Jones:1974pg}
\bibitem{Jones:1974pg}
  D.~R.~T.~Jones,
  %``Asymptotic Behavior of Supersymmetric Yang-Mills Theories in the Two Loop Approximation,''
  Nucl.\ Phys.\ B {\bf 87} (1975) 127.
  %doi:10.1016/0550-3213(75)90256-4
  %%CITATION = doi:10.1016/0550-3213(75)90256-4;%%

\end{thebibliography}
\end{document}